\documentclass[prd,preprint,
superscriptaddress,
nofootinbib,%
tightenlines
]{revtex4}
\usepackage{mathrsfs}
\usepackage{latexsym, bm}
\usepackage{graphicx}
\usepackage{indentfirst}
\usepackage{slashed}
\usepackage{amssymb}
\usepackage{amsmath}
\usepackage{bbm}
\usepackage{color}
\usepackage[dvipsnames]{xcolor}
\usepackage{epsfig}
\usepackage[titletoc]{appendix}
\usepackage{multirow}%
\usepackage{rotating}
\usepackage{epstopdf}
\usepackage{extarrows}
\usepackage{tabularx}
\usepackage{soul} 
\usepackage{xcolor}
\usepackage{lipsum}
\usepackage[colorlinks=true,allcolors=BlueViolet,hyperfootnotes=false]{hyperref}

\usepackage{physics}   

\allowdisplaybreaks
\begin{document}	
\title{\Large Low-energy elastic (anti)neutrino-nucleon scattering in covariant baryon chiral perturbation theory}
\author{Jin-Man~Chen}
\affiliation{School of Physics and Electronics, Hunan University, 410082 Changsha, China}
\author{Ze-Rui~Liang}
\affiliation{School of Physics and Electronics, Hunan University, 410082 Changsha, China}
\author{De-Liang Yao}
\email{yaodeliang@hnu.edu.cn}
\affiliation{School of Physics and Electronics, Hunan University, 410082 Changsha, China}
\affiliation{Hunan Provincial Key Laboratory of High-Energy Scale Physics and Applications, Hunan University, 410082 Changsha, China}
	
\date{\today} 
	
\begin{abstract}
The low-energy antineutrino- and neutrino-nucleon neutral current elastic scattering is studied within the framework of the relativistic SU(2) baryon chiral perturbation theory up to the order of $\mathcal{O}(p^3)$. We have derived the model-independent hadronic amplitudes and extracted the form factors from them. It is found that differential cross sections ${{\rm d} \sigma}/{{\rm d} Q^2}$ for the processes of (anti)neutrino-proton scattering are in good agreement with the existing MiniBooNE data in the $Q^2$ region $[0.13,0.20]$~GeV$^2$, where nuclear effects are expected to be negligible. For $Q^2\leq 0.13$~GeV$^2$, large deviation is observed, which is mainly owing to the sizeable Pauli blocking effect. Comparisons with the simulation data produced by the NuWro and GIENE Mento Carlo events generators are also discussed. The chiral results obtained in this work can be utilized as inputs in various nuclear models to achieve the goal of precise determination of the strangeness axial vector form factor, in particular when the low-energy MicroBooNE data are available in the near future.
\end{abstract}

\maketitle

\section{Introduction}

Neutrino-nucleon scattering, as one of the fundamental processes of neutrino interactions with matter, has been extensively investigated for decades for its importance in our understanding of a wide variety of physical phenomena; see e.g. Refs.~\cite{Formaggio:2012cpf,SajjadAthar:2021prg} for reviews. A comprehensive knowledge of the neutrino-nucleon/nuclei interactions is important to achieve the precision goal of modern neutrino-oscillation experiments~\cite{Super-Kamiokande:2010tar,MINOS:2011neo,DayaBay:2012fng,Benhar:2015wva,NuSTEC:2017hzk}. Furthermore, the neutral current scatterings play a dominant role in the thermal coupling between neutrinos and the stellar environment, thereby impacting the mean energies of various neutrino types in core-collapse supernovae~\cite{Bertulani:2009mf,Janka:2012wk}. Amongst them, a fundamental process is the $\nu N$ neutral current elastic (NCE) scattering. As nucleons are usually embedded in nuclei, a good and deep understanding of the $\nu N$ NCE interaction is indispensable for a precise extraction of nuclear effects~\cite{LlewellynSmith:1971uhs}, which is one of the main sources of systematical uncertainty in accurate determination of neutrino properties. On the other hand, it plays an extremely important role in unveiling the electroweak properties of the nucleon. Unlike the charged current quasi elastic (CCQE) $\nu N$ interaction, which only involves isovector weak current, the NCE reaction is sensitive to both the isovector and isoscalar week currents~\cite{Alvarez-Ruso:2016mbu}. Therefore, it provides the possibility to explore the strange quarks, which exist as sea quarks in nucleons, through their isoscalar contribution to the NCE interaction. 

Experimental efforts dedicated to the measurements of $({\bar{\nu}})\nu$N NCE scatterings have been undertaken since the middle of 1980s~\cite{Ahrens:1986xe,Perevalov:2009zz,MiniBooNE:2010xqw,MiniBooNE:2013dds,Ren:2022qop}. One of the goals of these experiments is to determine the strangeness contribution to the nucleon spin $\Delta s$~\cite{EuropeanMuon:1987isl}, which is associated with the strangeness axial form factors $G^s_A(Q^2)$ by $\Delta s=G^s_A(Q^2=0)$, with $Q^2$ being the momentum transfer squared between the initial and final nucleons. The E734 experiment at Brookhaven National Laboratory (BNL) measured the differential cross sections ${\rm d}\sigma/{\rm d}Q^2$ for (anti)neutrino-proton NCE scattering in the momentum transfer region $ 0.45 \lesssim Q^2  \lesssim 1.05 $~GeV$^2$, enabling them to analyze the strangeness contribution to axial vector form factor $G_A^s(Q^2)$ but with assumptions for its $Q^2$-behaviour~\cite{Ahrens:1986xe}. In 2010 and 2015, the MiniBooNE experiment at Fermi National Accelerator Laboratory released data of the differential cross-sections for neutrino- and antineutrino-induced NCE scattering in a mineral oil (CH$_2$) based target system~\cite{MiniBooNE:2010xqw,MiniBooNE:2013dds}, with a focus on the momentum transfer range below $2$ GeV$^2$. Note that the experimental data close to zero momentum transfer (i.e. $Q^2=0$) are provided in the Ph.D. dissertation by D.~Perevalov~\cite{Perevalov:2009zz}, a member of the MiniBooNE collaboration. It should be emphasized that the low $Q^2$ data of differential cross sections have a complicated varying behavior due to nuclear effects, which deserves more experimental investigations. To that end, the MicroBooNE experiment in Argon is conducting the measurements of muon neutrino NCE scattering from protons and more data will be available for $ 0.1 \lesssim Q^2 \lesssim 1 $~GeV$^2$~\cite{Ren:2022qop}. 

In the theoretical aspects, various phenomenological models are used to analyze the experimental data on $(\bar{\nu})\nu N$ NCE scattering, mainly focusing on the extraction of the neutral current axial form factors. For instance, with the dipole parametrization~\cite{Hand:1963zz}, the strange quark form factors of the proton and the axial vector dipole mass are determined by refitting the BNL experimental data on neutrino-proton elastic scattering~\cite{Garvey:1992cg}. In Ref.~\cite{Sufian:2018qtw}, Sufian {\it et al.} studied the NC weak axial form factor $G^Z_A(Q^2)$ by using the data from MiniBooNE experiments on neutrino-nucleon scattering and the result of the strange quark axial charge from lattice QCD~\cite{Liang:2018pis}. Therein, the model-dependent dipole parametrization is again employed for the $G^Z_A(Q^2)$. In a very recent work~\cite{Pate:2024acz}, the MiniBooNE NCE data are included in their global fit and an improvement is established in constraining the $G_A^s(Q^2)$. In addition to the dipole model, the $z$-expansion model is also chosen to parametrize the strangeness axial vector form factor. Here, we intend to carry out a model-independent calculation of the ($\bar{\nu}$)$\nu$-N NCE scattering within the framework of relativistic baryon chiral perturbation theory (ChPT). 

ChPT~\cite{Weinberg:1978kz, Gasser:1983yg, Gasser:1984gg} is the effective field theory of quantum chromodynamics (QCD) at low energies, and has been extensively used in hadron physics and nuclear physics; see~e.g. Refs.~\cite{Bernard:1995dp,Bernard:2007zu,Scherer:2002tk,Geng:2013xn,Yao:2020bxx} for reviews and applications. It was initially proposed for pure Goldstone bosons and later extended to describe interactions with nucleons~\cite{Gasser:1987rb}. As examples of applications to heavy hadrons, recent one-loop analyses of heavy charmed mesons and doubly charmed baryons in ChPT can be found e.g. in Refs.~\cite{Korpa:2022voo,Liang:2023scp}. Returning to the ${(\bar{\nu}})\nu$-$N$ scattering, systematical calculation in ChPT at one-loop order is rare. Low-energy theorems associated with neutrino induced one-pion production off the nucleon were derived using heavy baryon formalism in Ref.~\cite{Bernard:1993xh}. It is until 2018 that a first systematic study of charged-current weak pion production was carried out in relativistic baryon ChPT~\cite{Yao:2018pzc}. The study is later extended to the neutral-current weak pion production~\cite{Yao:2019avf}. 
The relevant formalism can also be applied to investigate the NCE ($\bar{\nu}$)$\nu$-$N$  scattering, which will provide useful information for the explanation of the existing MiniBooNE data and future MicroBooNE data in the low $Q^2$ regime.

In this work, the hadronic amplitude involved in the NCE $(\bar{\nu})\nu$-$N$ scattering is calculated in covariant baryon ChPT up to $\mathcal{O}(p^3)$. The ultraviolet (UV) divergences are dealt with by using the modified minimal subtraction ($\overline{\rm MS}$-$1$) scheme in dimensional regularization, while the notable power counting breaking (PCB) terms are removed by imposing the so-called extended-on-mass-shell (EOMS) scheme~\cite{Fuchs:2003qc}. In accordance with Lorentz and isospin symmetries, the amplitude can be decomposed and expressed in terms of $6$ structure functions: isovector (isoscalar) vector Dirac and Pauli form factors, isovector axial-vector and induced pseudoscalar form factors. Explicit expressions of the $6$ form factors are obtained and are consistent with the results derived in previous literature~\cite{Fuchs:2003ir,Yao:2017fym}. 

In our numerical computation, the low energy constants (LECs) are either determined elsewhere or set to be of natural size. We then predict the differential cross sections for four physical NCE processes: ${\nu}p\to {\nu} p$, $\bar{\nu}p\to \bar{\nu} p$, ${\nu} n\to {\nu} n$ and $\bar{\nu}n\to \bar{\nu} n$. It is found that our results of ${\rm d\sigma}/{\rm d}Q^2$ on ${\nu}p\to {\nu} p$ and $\bar{\nu}p\to \bar{\nu} p$ are in good agreement with the MiniBooNE data in the $Q^2$ region $\sim[0.13,0.20]$~GeV$^2$, where the ChPT calculation is expected to be reliable. Nevertheless, when $Q^2\leq 0.13$~GeV$^2$, deviation between our predictions and the experimental data can be seen. We ascribe the deviation to the occurrence of nuclear effects. It is well known that Pauli blocking is the dominant nuclear effect for small momentum transfer. It is then tested that, indeed, our ChPT results can be fine tuned by incorporating Pauli blocking effect, whose value is estimated by nuclear models implemented in NuWro Monte Carlo events generator~\cite{Juszczak:2005zs,Golan:2012rfa}.
For the NCE scattering on neutron, there are no experimental data so far. Therefore, we compare our ChPT results with the simulation data by NuWro, and large discrepancy is observed. Future experimental data for the neutron channel are appealed to explain this inconsistency, though the detection of neutron is challengeable. At last, total cross sections are also predicted and compared with the NuWro and GENIE~\cite{Andreopoulos:2009rq} data. The ChPT results of the NCE scattering on the nucleon, we obtain here, can be used as inputs in various nuclear models to extract the strangeness contribution to the nucleon spin and the strangeness axial vector form factor, especially when the MicroBooNE data appear in the near future.

The manuscript is organized as follows. Section~\ref{sec2.1} presents the basics of $({\bar{\nu}})\nu$-$N$ scattering, such as isospin and Lorentz decomposition. The calculation of the form factors in baryon ChPT is detailed in Section~\ref{sec.chpt}. In Section~\ref{sec3}, numerical results of differential and total cross sections are obtained, and comparisons to the MiniBooNE, NuWro and GIENE data are discussed. Section \ref{sec4} is our summary and outlook. Definition of the loop functions is given in Appendix~\ref{appA}, while the explicit expressions of hadronic amplitudes and form factors are relegated to Appendix~\ref{appB} and \ref{appC}, respectively.
	

\section{Basics of elastic neutrino-nucleon scattering}  \label{sec2.1}

\begin{figure}[tbp] 
\centering  
\includegraphics[width=6cm]{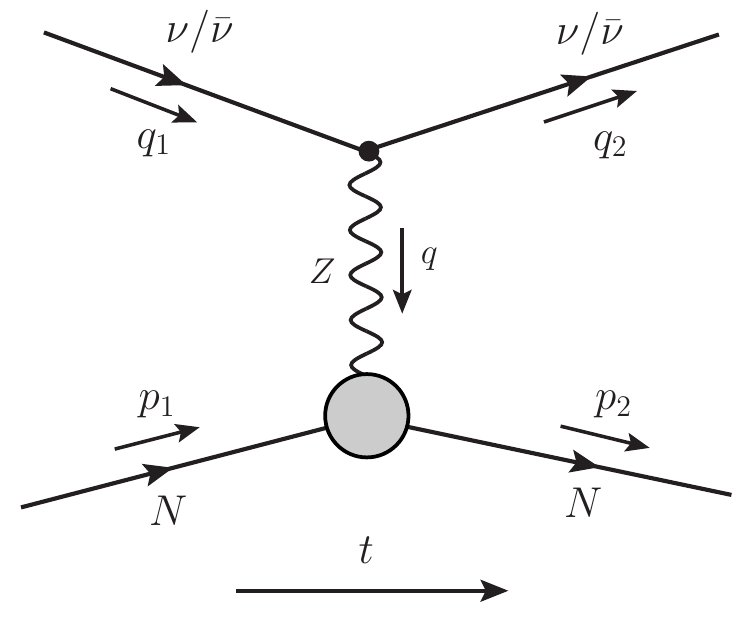}  
\caption{Kinematics of neutral current elastic neutrino-nucleon scattering under the one-boson exchange approximation.}  
\label{fig:nuN} 
\end{figure}

The process of $(\bar{\nu})\nu$-$N$ NCE scattering can be generally described by
\begin{align}
\ell(q_1) + N(p_1)\to \ell(q_2) + N(p_2) \ ,
\label{eq:NC} 
\end{align}
where $\ell=\nu $ or $\bar{\nu}$, $N=p$ or $n$. The four momenta of the incoming and outgoing particles are indicated in parentheses. The Lorentz-invariant Mandelstam variables are defined by
\begin{align}
	s=(p_1+q_1)^2 \ ,\quad
	t=(p_1-p_2)^2\ ,\quad
	u=(p_1-q_2)^2 \ ,
\end{align}
fulfilling the constraint $s+t+u=2m_N^2$, with $m_N$ being the physical mass of the nucleon. In the standard model (SM), this process is mediated by the vector $Z$ boson with momentum $q=p_2-p_1$, which is illustrated in Fig.~\ref{fig:nuN}.
Note that the one-boson exchange approximation is assumed. The energy region we are interested in is $ q^2 \ll M_Z^2$, where $ M_Z $ is the mass of $Z$-boson. Therefore, the scattering amplitude $ \mathcal{M} $ can be written as
\begin{align}
\mathcal{M}=-\frac{ G_F }{\sqrt{2}} L_\mu H^\mu~,
\end{align}
where $ G_F=1.166\times10^{-5} ~{\rm GeV}^{-2} $ is the Fermi constant~\cite{ParticleDataGroup:2022pth}; $L_\mu $ and $H^\mu$ are leptonic and hadronic matrix elements, respectively. Expressions of the leptonic and hadronic parts read
\begin{align}
L_\mu&= \bar{\nu}(q_2) \gamma_\mu (1-\gamma_5) \nu (q_1)  \ ,\notag\\
H^\mu&=\mel{N(p_2)}{\mathcal{J}_{NC}^\mu(0)}{N(p_1)}\ ,
\end{align}
where the neutral weak current is
\begin{align}
\mathcal{J}_{NC}^\mu =(1-2\sin^2\theta_W) \hat{\mathcal{V}}_3^\mu -\hat{\mathcal{A}}_3^\mu -2\sin^2\theta_W \hat{\mathcal{V}}_0^\mu\ .\label{eq.JNC}
\end{align}
with $\hat{\mathcal{V}}_3^\mu$, $\hat{\mathcal{A}}_3^\mu$ and $\hat{\mathcal{V}}_0^\mu$ being isovector vector, isovector axial-vector and isoscalar vector currents, in order. $\theta_W$ stands for the Weinberg angle. The leptonic part can be derived straightforwardly in the SM, while the hadronic one is complicated due to the inner structure of the nucleon. For the hadronic amplitude, one can perform isospin decomposition in the following way
\begin{align}
H^\mu=\chi^\dagger_f \left[\frac{\tau_a}{2} {H}^\mu_V +\frac{\tau_0}{2} {H}^\mu_S  \right] \chi_i  ~  , \quad a=3 ,
\end{align}
where $\chi_i $ and $\chi_f $ are isospin spinors of the initial and final nucleons, respectively. To be specific, one has $\chi_p=(1, 0)^T$ for the proton and $\chi_n=(0,1)^T$ for the neutron. ${H}^\mu_V $ and ${H}^\mu_S $ stand for the isospin vector and scalar amplitudes, respectively. In view of Eq.~\eqref{eq.JNC}, one readily has
\begin{equation}
\begin{aligned}
{H}^\mu_V &=(1-2\sin^2{\theta_W}) {V}^\mu_V-{A}^\mu_V ~,  \\
{H}^\mu_S &=-2\sin^2{\theta_W} {V}^\mu_S ~,
\end{aligned}
\end{equation}
with the matrix elements ${V}^\mu_{V,S}$ and ${A}^\mu_V$ expressed in terms of form factors~\cite{Bernard:2001rs,Alvarez-Ruso:2016mbu}
\begin{equation}
\begin{aligned}
{V}^\mu_V =& \bar{\mathbf{u}}(p_2) \Big[ \gamma^\mu F_1^V +\frac{i}{2m_N}\sigma^{\mu\nu}q_\nu F_2^V   \Big] \mathbf{u}(p_1) , \\
A^\mu_V =& \bar{\mathbf{u}}(p_2) \Big[  \gamma^\mu \gamma_5 {G}_A + \frac{q^\mu}{m}\gamma_5 {G}_P    \Big] \mathbf{u}(p_1)  , \\
{V}^\mu_S = &  \bar{\mathbf{u}}(p_2) \Big[ \gamma^\mu F_1^S +\frac{i}{2m_N}\sigma^{\mu\nu}q_\nu F_2^S   \Big] \mathbf{u}(p_1)  \ .
\label{eq:vas}
\end{aligned}
\end{equation}
In above, $\bar{\mathbf{u}}(p_2)$ and $\mathbf{u}(p_1)$ represent the spinors of the initial and final nucleons, respectively. There are six unknown form factors: ${F}_1^V$, ${F}_2^V$, ${G}_A$, ${G}_P$, ${F}_1^S$, ${F}_2^S$. 
${F}_1^V$ (${F}_1^S$) and ${F}_2^V$ (${F}_2^S$) are called isovector (isoscalar) Dirac and Pauli form factors, respectively. $G_A$ and $G_P$ are axial and induced pseudoscalar form factors of the nucleon, respectively. Both $G_A$ and $G_P$ are isovector, due to the absence of isosinglet axial current. We will calculate the form factors in the framework of covariant baryon ChPT in Section~\ref{sec.chpt}. 

In practice, it is often useful to reorganize the hadronic amplitude according to its Lorentz structure. That is, 
\begin{align}
H^\mu &\equiv \bar{\bf u}(p_2)\mathcal{H}^\mu{\bf u}(p_1)\ ,
\label{eq.GammaExp}\\
\mathcal{H}^\mu&=\gamma^\mu \mathcal{F}_1+\frac{i}{2m_N}\sigma^{\mu\nu}q_\nu \mathcal{F}_2-\gamma^\mu\gamma_5\mathcal{G}_A-\frac{q^\mu}{m_N}\gamma_5\mathcal{G}_P
\ ,
\end{align}
where the combined form factors $\mathcal{F}_{1,2}$ and $\mathcal{G}_{A,P}$ are related to the ones in Eq.~\eqref{eq:vas} via
\begin{align}
\mathcal{F}_i(t) &=  \cos{2\theta_W} {F}_i^V(t)  \frac{\mathcal{C}_{3}}{2} -  2  \sin^2{\theta_W} {F}_i^S(t) \frac{ \mathcal{C}_{0} }{2}   ~, \quad i=1,2 , \label{eq:FFsNC1}\\
\mathcal{G}_j(t) &=  {G}_j(t) \frac{\mathcal{C}_3}{2}~, \quad j=A,P ,
\label{eq:FFsNC2}
\end{align}
For brevity, isospin factors have been defined, i.e., $\mathcal{C}_{3}=\chi_f^\dagger \tau_3\chi_i$ and $\mathcal{C}_{0}=\chi_f^\dagger \tau_0\chi_i$. In this way, the hadronic amplitude for a given physical process can be obtained by properly choosing the values of $\mathcal{C}_{3}$ and $\mathcal{C}_{0}$. We show the isospin factors in Table~\ref{tab.iso} for easy reference.

\begin{table}[htb]
\centering
\caption{Isospin factors for physical processes.}
\label{tab.iso}
\begin{tabular}{c|cc}
\hline \hline 
physical process &  $\mathcal{C}_3$ & $\mathcal{C}_0$ \\
\hline
$\nu + p \to \nu + p $ &  ~~~~$1$~~~~ & ~~~~$1$~~~~ \\
$\bar{\nu} + p \to \bar{\nu} + p $ &   $1$ & $1$ \\
$\nu + n \to \nu + n $ &  $-1$ & $1$ \\
$\bar{\nu} + n \to \bar{\nu} + n $  & $-1$ & $1$ \\
\hline \hline
\end{tabular}
\end{table}

\section{Hadronic form factors in baryon ChPT}
\label{sec.chpt}

The hadronic form factors encode dynamical information of strong interaction inside the nucleon. At low energies, the fundamental theory quantum chromodynamics (QCD) is not feasible anymore due to the non-perturbative nature of strong interaction. Instead, a popular and powerful tool for the study of the low-energy dynamics is ChPT, which is an effective field theory of QCD in this regime. In this section, we are going to calculate the form factors involved in the $(\bar{\nu})\nu$-$N$ scattering within the framework of relativistic baryon ChPT up to $\mathcal{O}(p^3)$ by using EOMS scheme.

\subsection{Chiral effective Lagrangian}

The chiral effective Lagrangian pertinent to our calculation can be organized in the form as
\begin{equation}
\mathcal{L}_{\rm eff}=\mathcal{L}_{\pi N}^{(1)}+\mathcal{L}_{\pi N}^{(2)}+\mathcal{L}_{\pi N}^{(3)} 
+\mathcal{L}_{\pi \pi }^{(2)} + \mathcal{L}_{\pi \pi }^{(4)}\ ,
\end{equation}	
where the superscripts denote the chiral orders. The pieces $\mathcal{L}_{\pi \pi}^{(i)}$ ($i=2,4$) describe the purely pionic interactions, while the terms $\mathcal{L}_{\pi N}^{(j)}$ ($j=1,2,3$) are constructed by further including the nucleon fields as explicit degrees of freedom.

The $\pi\pi$ Lagrangians are given by~\cite{Gasser:1983yg}
\begin{align}
\mathcal{L}_{\pi \pi}^{(2)} &= \frac{F^2}{4} \langle u_\mu u^\mu + \chi_+ \rangle \ ,  \\
\mathcal{L}_{\pi \pi}^{(4)} &=\frac{1}{8} \ell_4 \langle u_\mu u^\mu \rangle \langle \chi_+ \rangle + \frac{1}{16} ( \ell_3 + \ell_4 ) \langle \chi_+ \rangle^2\ ,
\end{align}
where $F$ is the pion decay constant in the chiral limit and $\ell_i(i=3,4)$ are unknown LECs. The chiral operators in the $\mathcal{O}(p^4)$ Lagrangian are required for the renormalization of the pion mass and decay constant. The so-called chiral vielbein reads
\begin{align}
    u_\mu =i\left\{ u^\dagger(\partial_\mu - ir_\mu)u - u(\partial_\mu - i l_\mu)u^\dagger \right\}  ,
\end{align}
and the Goldstone pion fields are collected in the unitary $ 2\times 2$ matrix $ u $,
\begin{equation}
		u=\exp\left(i\frac{\Phi}{\sqrt{2}F}\right)  \ ,\quad\Phi =\pi^a \tau^a  =\begin{pmatrix} \pi^0 & \sqrt{2} \pi ^+ \\  \sqrt{2} \pi^- & -\pi^0  \end{pmatrix} 
\end{equation}
with $\tau^a$ the Pauli matrices. Furthermore, $l_\mu$ and $r_\mu$ are left- and right-handed external fields. The chiral blocks $\chi_{\pm}$ are defined by
\begin{align}
\chi_\pm &=u^\dagger \chi u^\dagger \pm u \chi^\dagger u \ , \quad
\chi ={\rm diag}(M^2,M^2) \ ,
\end{align}
where $ M $ is the pion mass in the SU(2) chiral limit.

The leading order $ \pi N $ effective Lagrangian reads~\cite{Fettes:2000gb}
\begin{equation}
\mathcal{L}_{\pi N}^{(1)}=\bar{\Psi}_N \big\{ i\slashed{D}-m+\frac{g}{2}\slashed{u}\gamma_5\big\} \Psi_N  ,
\label{eq:LpiN1}
\end{equation}
with $m$ and $g$ being the nucleon mass and axial coupling constant in the chiral limit, respectively. The isodoublet $\Psi_N$ comprises the proton and neutron fields, i.e., $\Psi_N=(p,n)^T$. The covariant derivative acting on the nucleon fields is given by
\begin{align}
D_\mu &= \partial_\mu +\Gamma_\mu -i \hat{v}_\mu^s ~,\\
\Gamma_\mu &=\frac{1}{2}\left\{ u^\dagger(\partial_\mu - ir_\mu)u + u(\partial_\mu - i l_\mu)u^\dagger \right\} \ .
\end{align} 
Here, $\hat{v}^s_\mu$ denotes the isosinglet vector field.

The full set of $\pi N$ Lagrangians at $\mathcal{O}(p^2)$ and $\mathcal{O}(p^3)$ can be found, e.g., in Ref.~\cite{Fettes:2000gb}. Here, we only display the terms relevant to our calculation:
\begin{equation}
\begin{aligned}
\mathcal{L}_{\pi N}^{(2)}= \frac{1}{8m} c_6 \bar{\Psi}_N \Big\{ F^+_{\mu \nu} \sigma^{\mu \nu} \Big\}\Psi_N  + \frac{1}{8m} c_7 \bar{\Psi}_N\Big\{  \langle F^+_{\mu \nu}\rangle \sigma^{\mu \nu} \Big\}\Psi_N +\cdots ,
\end{aligned}
\end{equation}
\begin{equation}
\begin{aligned}
\mathcal{L}_{\pi N}^{(3)}&= \bar{\Psi}_N\Big\{  \frac{i}{2m} d_6[D^\mu,\widetilde{F}^+_{\mu\nu} ] D^\nu +h.c. \Big\}\Psi_N + \bar{\Psi}_N\Big\{  \frac{i}{2m} d_7[D^\mu, \langle {F}^+_{\mu\nu} \rangle ] D^\nu  +h.c. \Big\}\Psi_N \\
&\quad+ \bar{\Psi}_N\Big\{ \frac{1}{2}d_{16}\gamma^\mu \gamma_5 \langle \chi_+ \rangle  u_\mu \Big\}\Psi_N   +\bar{\Psi}_N\Big\{ \frac{i}{2}d_{18}\gamma^\mu \gamma_5 [ D_\mu , \chi_- ] \Big\}\Psi_N \\
&\quad+ \bar{\Psi}_N\Big\{ \frac{1}{2}d_{22}\gamma^\mu \gamma_5 [ D^\nu , F^-_{\mu \nu} ] \Big\}\Psi_N +\cdots~,
\end{aligned}
\end{equation}
where $ c_i~ (i=6,7) $ and $ d_j~  (j=6,7,16,18,22) $ are unknown LECs, and $h.c.$ refers to the hermitian conjugation. The field tensors, regarding the $l_\mu$, $r_\mu$ and $\hat{v}_\mu$ external sources, are defined as follows: 
\begin{align}
F^+_{\mu\nu}&=f^+_{\mu\nu} + 2\hat{v}_{\mu\nu}^{(s)} \ ,\\
F^-_{\mu\nu} &=f^-_{\mu\nu}\ ,\\
f^\pm_{\mu\nu} &=u f^L_{\mu\nu}u^\dagger \pm u^\dagger f^R_{\mu\nu}u , \\
\hat{v}_{\mu\nu}^{(s)}&= \partial_\mu \hat{v}^{(s)}_\nu - \partial_\nu \hat{v}^{(s)}_\mu \ ,
\end{align} 
with the left- and right-handed external field strength tensors given by
\begin{align}
f^L_{\mu\nu}&=\partial_\mu l_\nu -\partial_\nu l_\mu - i[ l_\mu , l_\nu ]  ,\\
f^R_{\mu\nu}&=\partial_\mu r_\nu -\partial_\nu r_\mu - i[ r_\mu , r_\nu ] \ .
\end{align} 
The traceless tensor $\widetilde{F}^+_{\mu\nu}$ can be obtained through
\begin{align}
\widetilde{F}^+_{\mu\nu}=F^+_{\mu\nu}-\frac{1}{2} \langle F^+_{\mu\nu} \rangle \ ,    
\end{align}
where $\langle\cdots\rangle$ denotes the trace in the flavor space. The coupling of the $Z$ boson is incorporated by setting	
\begin{align}
l_\mu &=\left( \frac{g_W}{2\cos{\theta_W}} \right)  \left( -2\cos^2{\theta_W} \right) \mathcal{Z}_\mu \frac{\tau_3}{2} , \\
r_\mu &=\left( \frac{g_W}{2\cos{\theta_W}} \right)  \left( 2\sin^2{\theta_W} \right) \mathcal{Z}_\mu \frac{\tau_3}{2}  , \\
\hat{v}_\mu^{(s)}&=\left( \frac{g_W}{2\cos{\theta_W}} \right)  \left( 2\sin^2{\theta_W} \right) \mathcal{Z}_\mu \frac{\tau_0}{2} ,  \label{eq:vmus} 
\end{align} 
with $g_W$ the weak coupling constant. The factor ${g_W}/{(2\cos{\theta_W})}$ is extracted out from the hadronic part to define the Fermi constant $G_F$ in Eq.~\eqref{eq.GammaExp}, i.e., $G_F={\sqrt{2}g_W^2}/({8M_Z^2\cos^2\theta_W})$.

\subsection{Form factors}

\begin{figure*}[htbp]  
\includegraphics[width=0.75\textwidth]{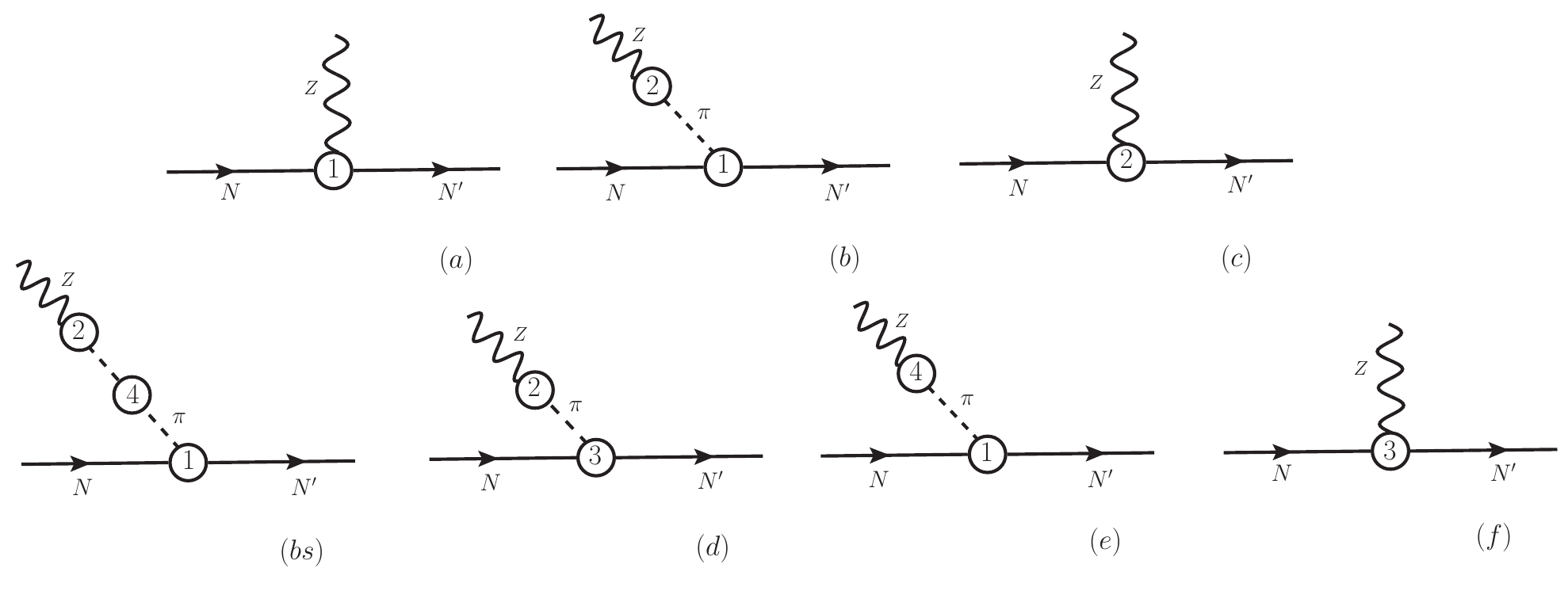}  
\caption{Tree-level Feynman diagrams up to and including $\mathcal{O}(p^3)$. The solid, dashed and wavy lines represent the nucleon, pions and the $Z$ boson, in order. The circled numbers indicate the chiral orders of the vertices.}
\label{fig:nuNtree}
\end{figure*}

\begin{figure}[htbp]  
\centering  
\includegraphics[width=0.9\textwidth]{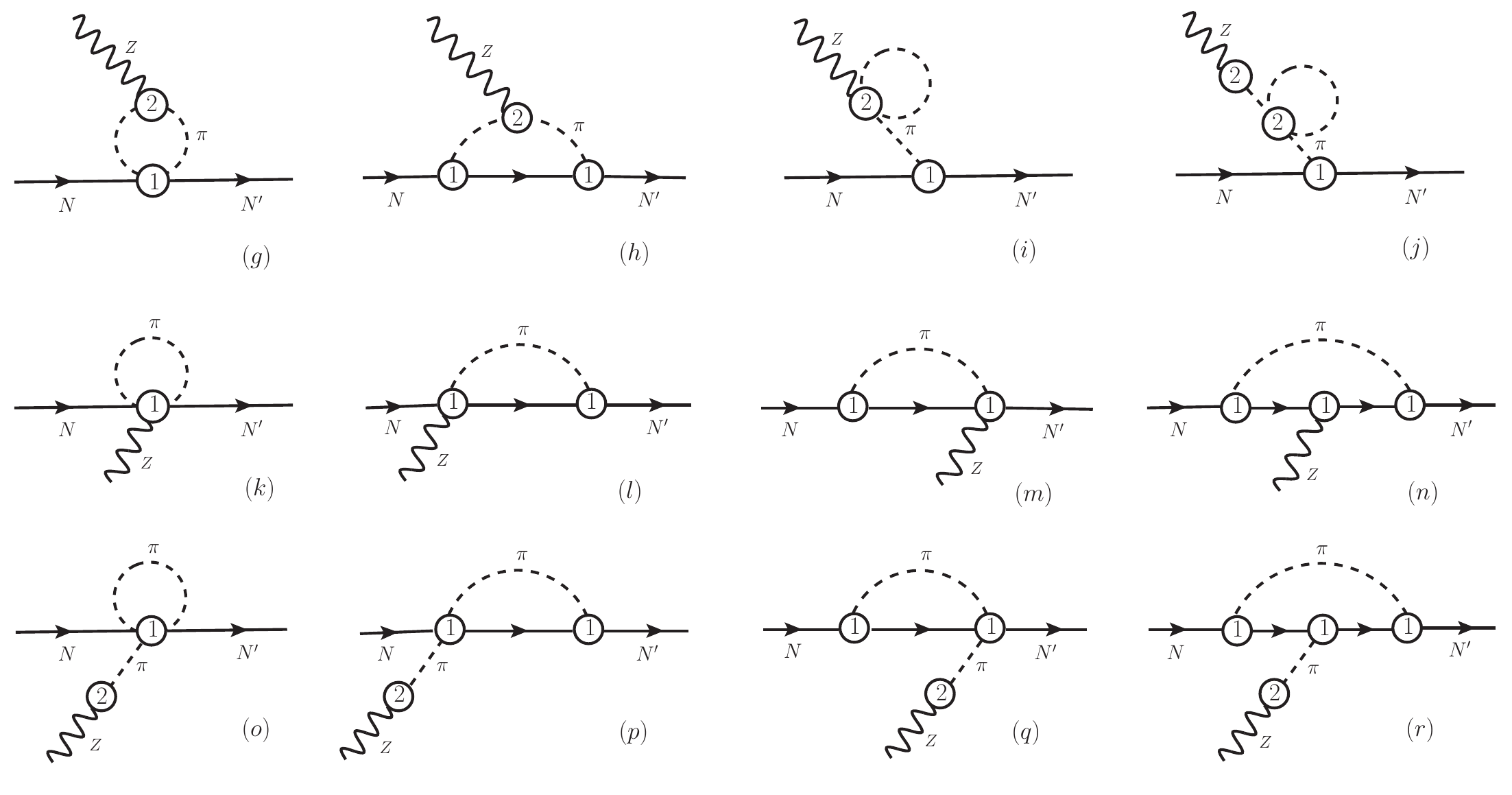} 
\caption{One-loop Feynman diagrams up to and including $\mathcal{O}(p^3)$. The solid, dashed and wavy lines represent the nucleon, pions and the $Z$ boson, in order. The circled numbers indicate the chiral orders of the vertices.}
\label{fig:nuNloop} 
\end{figure}

With the Lagrangians specified in the previous subsection, we are now in the position to calculate the form factors up to and including $\mathcal{O}(p^3)$. Tree and leading one-loop Feynman diagrams are shown in Fig.~\ref{fig:nuNtree} and Fig.~\ref{fig:nuNloop}, respectively. Diagrams with one-loop corrections on the external nucleon lines are not displayed explicitly, which will be taken into account by the wave function renormalization procedure. 

The obtained hadronic amplitudes are shown diagram by diagram in Appendix~\ref{appB}, from which one can extract the form factors. The results of form factors are expressed as
\begin{align}
{F}^V_1(t)=&1-2d_6t+F_1^{V,{\rm loops}}+F_1^{V,{\rm wf}}\ ,\notag \\
{F}^V_2(t)=& c_6+2d_6 t+{F}_2^{V,{\rm loops}}+F_2^{V,{\rm wf}}\ , \notag\\
{F}^S_1(t)=&1-4d_7t+F_1^{S,{\rm loops}}+F_1^{S,{\rm wf}}\ ,\notag \\
{F}^S_2(t)=& (c_6+2c_7)+4d_7 t+{F}_2^{S,{\rm loops}}+{F}_2^{S,{\rm wf}}\ ,\notag \\
{G}_A(t)=&g+(4d_{16}M^2+d_{22}t)+G_A^{\rm loops}+G_A^{\rm wf}\ ,\notag\\
G_P(t) =&\frac{2 g  m_N^2}{M^2-t}+\frac{4m_N^2 M^2 (2d_{16}-d_{18})}{M^2-t}+\frac{4g m_N^2 M^2 \ell_4}{F^2(M^2-t)}- 2m_N^2 d_{22} \notag \\
&  -\frac{4 g m_N^2 M^2 \big[ M^2 \ell_3 + (M^2-t) \ell_4 \big]}{F^2 (M^2-t)^2} +G_P^{\rm loops}+G_P^{\rm wf}\ ,
\label{eq.expressions}
\end{align}
where the expressions of loop contributions are relegated to Appendix~\ref{appC}. Note that the induced pseudoscalar form factor does not contribute to the cross sections if the neutrino mass is zero, but we show it here for completeness. The last terms in the above equations account for the effect of wave function renormalization. Up to $\mathcal{O}(p^3)$, they are given by 
\begin{align}
T^{\rm wf} = (T^{(a)}+T^{(b)})(\mathcal{Z}_N-1)+ \cdots\ ,\quad T\in\{F_1^V, F_2^V, F_1^S, F_2^S, G_A, G_P\}\ ,
\end{align}
where the ellipsis stands for the higher order terms beyond our accuracy. $\mathcal{Z}_N$ is the wave function renormalization constant of the nucleon at leading one-loop order
\begin{align}
\mathcal{Z}_N=1+\delta^{(2)}_{\mathcal{Z}_N}+\mathcal{O}(p^3)
\end{align}
where the $\mathcal{O}(p^2)$ term reads~\cite{Chen:2012nx}
\begin{align}
\delta^{(2)}_{\mathcal{Z}_N}=\frac{3g^2}{4F_\pi^2(4m_N^2-m_\pi^2)}\big\{&4m_\pi^2 [A_0(m_N^2)+(m_\pi^2-3m_N^2)B_0(m_N^2,m_\pi^2,m_N^2)-m_N^2]\notag\\
&+(12m_N^2-5m_\pi^2)A_0(m_\pi^2)\big\}\ .
\end{align}
For the definitions of the loop integrals $A_0$ and $B_0$, the readers are referred to Appendix~\ref{appA}. Note that the bare parameters $F$, $m$ and $M$ have been replaced by the corresponding physical ones $F_\pi$, $m_N$ and $m_\pi$, since the resultant difference is of higher order beyond our accuracy. 

The loop amplitudes are calculated by imposing the dimensional regularization, and the UV divergence in the loop integrals are subtracted with the so-called $\overline{\rm MS}-1$ subtraction scheme. Actually, the UV divergence are cancelled out by the counter terms from the chiral effective Lagrangian. In practice, one splits the bare LECs into two parts: a finite piece and a divergent one. To be specific, the LECs showing up in Eq.~\eqref{eq.expressions} are separated in the following manner
\begin{equation}
X=X^r + \dfrac{\beta_X}{16\pi^2}R , ~~ X \in \{g, c_{6,7}, d_{6,7,16,18,22}, \ell_{3,4} \}  ,
\end{equation}
where $ R=2/(d-4)+\gamma_E-1-\ln(4\pi) $, $d$ is the dimension of space-time, $ \gamma_E $ denotes Euler constant, and $ \beta_X $ is called beta functions. In order to remove the UV divergence from the loops, the $ \beta_X $'s have to take the following values,
\begin{equation}
\begin{aligned}     
&\beta_g=\frac{g(g^2-2)m^2}{F^2}\ ,\quad \beta_{c_6}=\beta_{c_7}=0\ , \quad \beta_{d_6}=\frac{g^2-1}{12F^2}\ ,\quad \beta_{d_{16}}=\frac{g(g^2-1)}{4F^2} , \\
&\beta_{d_7}=\beta_{d_{18}}=\beta_{d_{22}}=0 , \quad \beta_{\ell_3}=-\frac{1}{4} , \quad \beta_{\ell_4}=1 .
\end{aligned}
\label{eq.UVcan}
\end{equation}
	
It is pointed out in Ref.~\cite{Gasser:1987rb} that the PCB issue arises when the nucleons appear as internal lines in the loop diagrams. To restore power counting, here we employ the EOMS scheme, in which an extra finite renormalization is required. In our case, the PCB terms are those pieces, stemming from the loops, with chiral orders lower than $\mathcal{O}(p^3)$. As discussed in Ref.~\cite{Fuchs:2003qc}, those terms are polynomials of, e.g., pion masses, which are called infrared regular terms therein, and hence can be absorbed by shifting the LECs of the tree amplitudes. That is, in addition to Eq.~\eqref{eq.UVcan}, the UV-finite $\mathcal{O}(p)$ and $\mathcal{O}(p^2)$ LECs are further split, 
\begin{equation}
X^r=\tilde{X} + \dfrac{ \tilde{\beta}_X}{16\pi^2}R , ~~ X \in \{ g, c_{6,7} \}  ,
\end{equation}
with the beta functions $ \tilde{\beta}_X $'s  given by
\begin{equation}
\tilde{\beta}_{g}=-\frac{g(g^2-2)\bar{A}_0(m^2)}{F^2} , ~~ \tilde{\beta}_{c_6}=-\frac{5g^2m^2}{F^2} , ~~ \tilde{\beta}_{c_7}=\frac{4g^2m^2}{F^2} .
\end{equation}
Here the function $\bar{A}_0(m^2)$ represents the UV-subtracted one-point loop integral. Namely, $\bar{A}_0(m^2)$ is obtained by removing the piece  proportional to $R$ in Eq.~\eqref{eq.A0}.
	
\section{Numerical Results and Discussions}  \label{sec3}
	
\subsection{Differential Cross Section}
In the laboratory frame, the differential cross section with respect to the momentum transfer squared reads
\begin{align}
\frac{{\rm d}\sigma}{{\rm d}Q^2} = \frac{ \overline{|\mathcal{M}|}^2}{64\pi m_N^2 E_\nu^2}~,
\label{eq:dcs1}
\end{align}
where $ Q^2\equiv -q^2 =-t $. $ E_\nu$ is the neutrino energy, and $ m_N  $ is the physical nucleon mass. The spin-averaged amplitude squared can be written as 
\begin{align}
\overline{|\mathcal{M}|}^2=\frac{G_F^2}{2} L_{\mu \nu} H^{\mu \nu} \ ,
\label{eq:M^2}
\end{align}
with the leptonic tensor
\begin{align}
L_{\mu \nu}=8\Big[ q_{1\mu}q_{2\nu}-g_{\mu \nu}q_{1} \cdot q_{2}+q_{1\nu}q_{2\mu} \pm i\epsilon _{\mu  \lambda\nu\sigma }q_{1}^{\lambda}q_{2}^{\sigma} \Big] \ ,
\label{eq:L_ten}
\end{align}
where the \lq\lq$+$" and \lq\lq$-$" signs correspond to the neutrino and antineutrino cases, respectively. The hadronic tensor is given by
\begin{align}
H_{\mu \nu}=\frac{1}{2}{\rm Tr}\Big[ \big( \slashed{p}_1 + m_N \big) \widetilde{\mathcal{H}}_\mu \big( \slashed{p}_2 + m_N \big) \mathcal{H}_\nu \Big]  ,
\label{eq:H_ten}
\end{align}
where $\widetilde{\mathcal{H}}_\mu = \gamma_0 \mathcal{H}_\mu^\dagger \gamma_0 $ and $\mathcal{H}_\mu$ is defined in Eq.~\eqref{eq.GammaExp}. One can further contract the leptonic and hadronic tensors, which leads to the traditional form of the differential cross section~\cite{LlewellynSmith:1971uhs,Garvey:1992cg}:
\begin{gather}
\frac{{\rm d}\sigma}{\mathrm{d}{Q^2}}= \frac{G_F^2 m_N^2 }{8\pi E_\nu^2} \Big[A(Q^2)\pm \frac{(s-u)}{m_N^2} B(Q^2)+ \frac{(s-u)^2}{m_N^4} C(Q^2) ~\Big] \ ,
\label{eq:dcs2}
\end{gather}
where the scalar functions $ A(Q^2) $, $ B(Q^2) $, and $ C(Q^2) $ are related to the form factors in Eq.~\eqref{eq:FFsNC1} and Eq.~\eqref{eq:FFsNC2} through
\begin{equation}
\begin{aligned}
A(Q^2)&\equiv 4\eta \Big[ \mathcal{G}_A^2(Q^2) \big( 1+\eta \big)+4\eta \mathcal{F}_1(Q^2) \mathcal{F}_2(Q^2)   \\
&\qquad \quad -\Big( \mathcal{F}_1^2(Q^2)-\eta \mathcal{F}_2^2(Q^2) \Big) \big( 1-\eta \big) \Big] ~,\\
B(Q^2)&\equiv 4\eta~ \mathcal{G}_A(Q^2) \Big(\mathcal{F}_1(Q^2)+\mathcal{F}_2(Q^2) \Big) ~,\\
C(Q^2)&\equiv \frac{1}{4} \Big[ \mathcal{G}^2_A(Q^2) + \mathcal{F}^2_1(Q^2)+\eta \mathcal{F}^2_2(Q^2) \Big] ~,
\label{eq:ABC}
\end{aligned}
\end{equation}
with $(s-u)=4m_N E_\nu-Q^2 $ and $\eta =Q^2/4m_N^2$. It should be noted that the neutrino mass is zero and consequently the pseudoscalar form factor $G_P$ is absent. With the above formulae, we are in the position to compute the cross sections for physical processes numerically.

\begin{table}[htp]
\centering
\caption{Values of the LECs involved in neutrino-nucleon NCE scattering. }
\begin{tabular}{>{\centering\arraybackslash}m{1.5cm}  >{\centering\arraybackslash}m{1.5cm}  >{\centering\arraybackslash}m{3cm}  >{\centering\arraybackslash}m{5cm} } 
\hline \hline
&  LEC  &   Value  &   Source   \\
\hline
\vspace{0.1cm}
$\mathcal{L}_{\pi N}^{(1)}$ & $g$ &$  1.13\pm 0.01 $ &$G_A$~\cite{Yao:2017fym}\\
$\mathcal{L}_{\pi N}^{(2)}$ &  $ c_6 $ &  $1.35 \pm 0.04$  &  $\mu_p$ and $\mu_n$ \cite{Bauer:2012pv,patrignani2016review}   \\
&  $ c_7 $ &  $-2.68 \pm 0.08$   &  $\mu_p$ and $\mu_n$ \cite{Bauer:2012pv,patrignani2016review}   \\
$\mathcal{L}_{\pi N}^{(3)}$  &  $ d_6 $ &  $0.0\pm 1.0$  &   -  \\
&  $ d_7 $ &  $-0.49$  &  electromagnetic radii~\cite{Fuchs:2003ir}   \\
&  $ d_{16} $ &  $-0.83 \pm 0.03$  &   $G_A$~\cite{Yao:2017fym}   \\
&  $ d_{22} $ &   $0.96 \pm 0.03$   &  $G_A$~\cite{Yao:2017fym}    \\
\hline \hline
\end{tabular}
\label{tab:LECs}
\end{table}

In our numerical computation, we use $m_N=938.8$~MeV, $m_\pi=138$~MeV and $F_\pi=92.21$~MeV~\cite{ParticleDataGroup:2022pth}. Values of the LECs relevant to NCE scattering are summarized in Table~\ref{tab:LECs}. The values of $g$, $d_{16}$ and $d_{22}$ are taken from Ref.~\cite{Yao:2017fym}. They appear in the axial form factor $G_A$ and were pinned down by fitting to the corresponding lattice QCD data simulated at unphysical pion masses~\cite{Yao:2017fym}. The LECs $c_6$ and $c_7$ have been determined by employing the empirical values of the anomalous magnetic moments of the nucleon ($\kappa_p$ and $\kappa_n$) in Ref.~\cite{Bauer:2012pv,patrignani2016review}. We set $d_7=-0.49$, which is extracted from the neutron and proton electromagnetic radii in Ref.~\cite{Fuchs:2003ir}. The $d_6$ is set to be of natural size, following Ref.~\cite{Yao:2018pzc}. The renomalization scale $\mu$ in the loop integrals is fixed at the physical nucleon mass: $\mu=m_N$. We do not assign values to the LECs only showing up in $G_P$, since they are irrelevant to the NCE scattering cross sections. 

\begin{figure}[htbp]	
\includegraphics[width=0.75\textwidth]{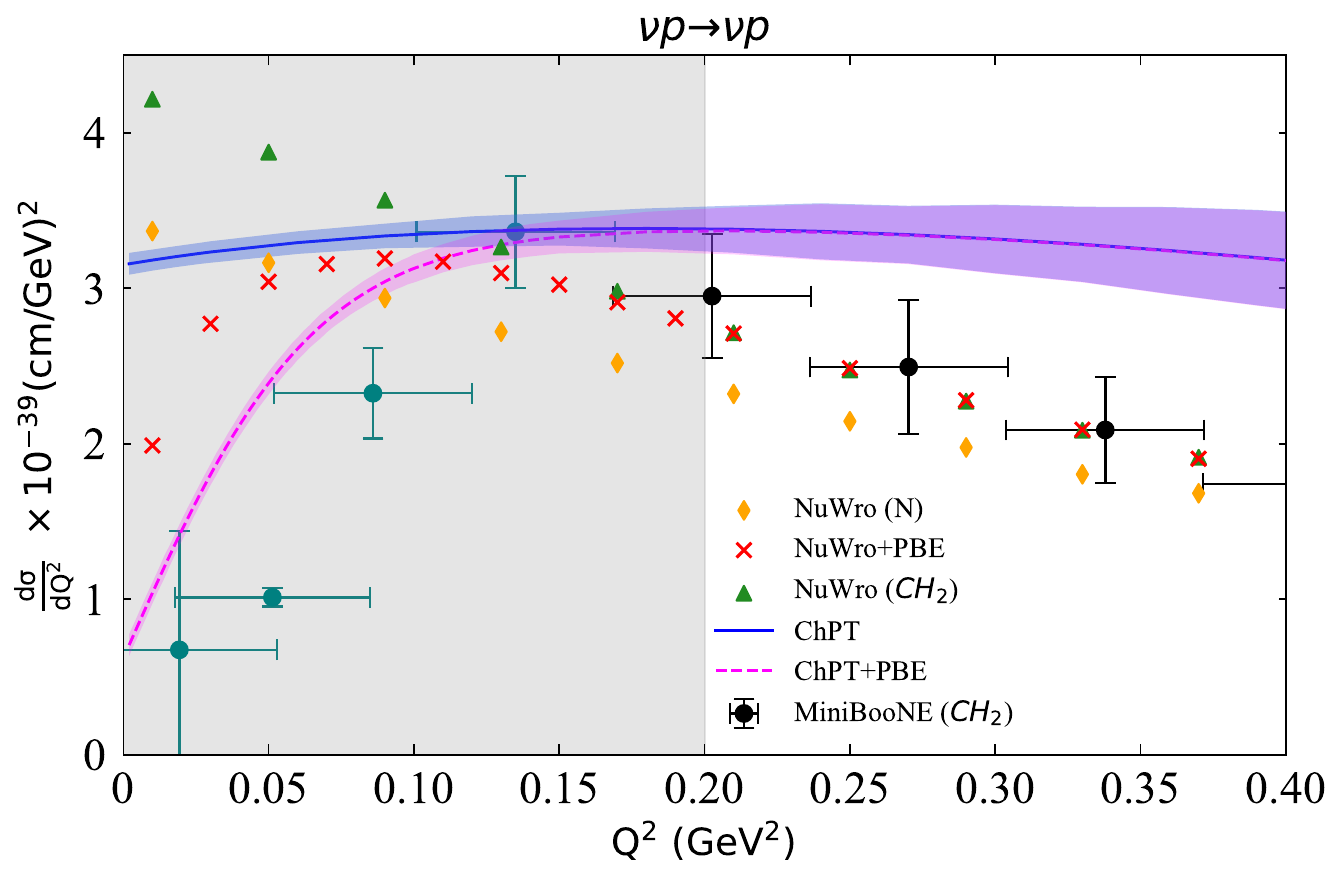} 
\caption{Differential cross section ${\rm d\sigma}/{\rm d }Q^2$ of $\nu$-$p$ NCE scattering, with the neutrino energy being $E_\nu=0.8$~GeV. The MiniBooNE data~\cite{MiniBooNE:2013dds,MiniBooNE:2010xqw} are marked by black dots with error bars. The NuWro data for free nucleon, bound nucleon with and without Pauli blocking effect are represented by orange diamonds, red crosses and green triangles in order. The blue solid line denotes the ChPT result up to $\mathcal{O}(p^3)$, while the magenta dashed line stands for the sum of our ChPT prediction and Pauli blocking effect. The error bands are obtained by varying the LECs in their $1$-$\sigma$ uncertainties. We use the abbreviation \lq\lq PBE" for \lq\lq Pauli blocking effect" in the figure.}  
\label{fig:nproton_dcs_band}
\end{figure}

\begin{figure}[tp]
\includegraphics[width=0.75\textwidth]{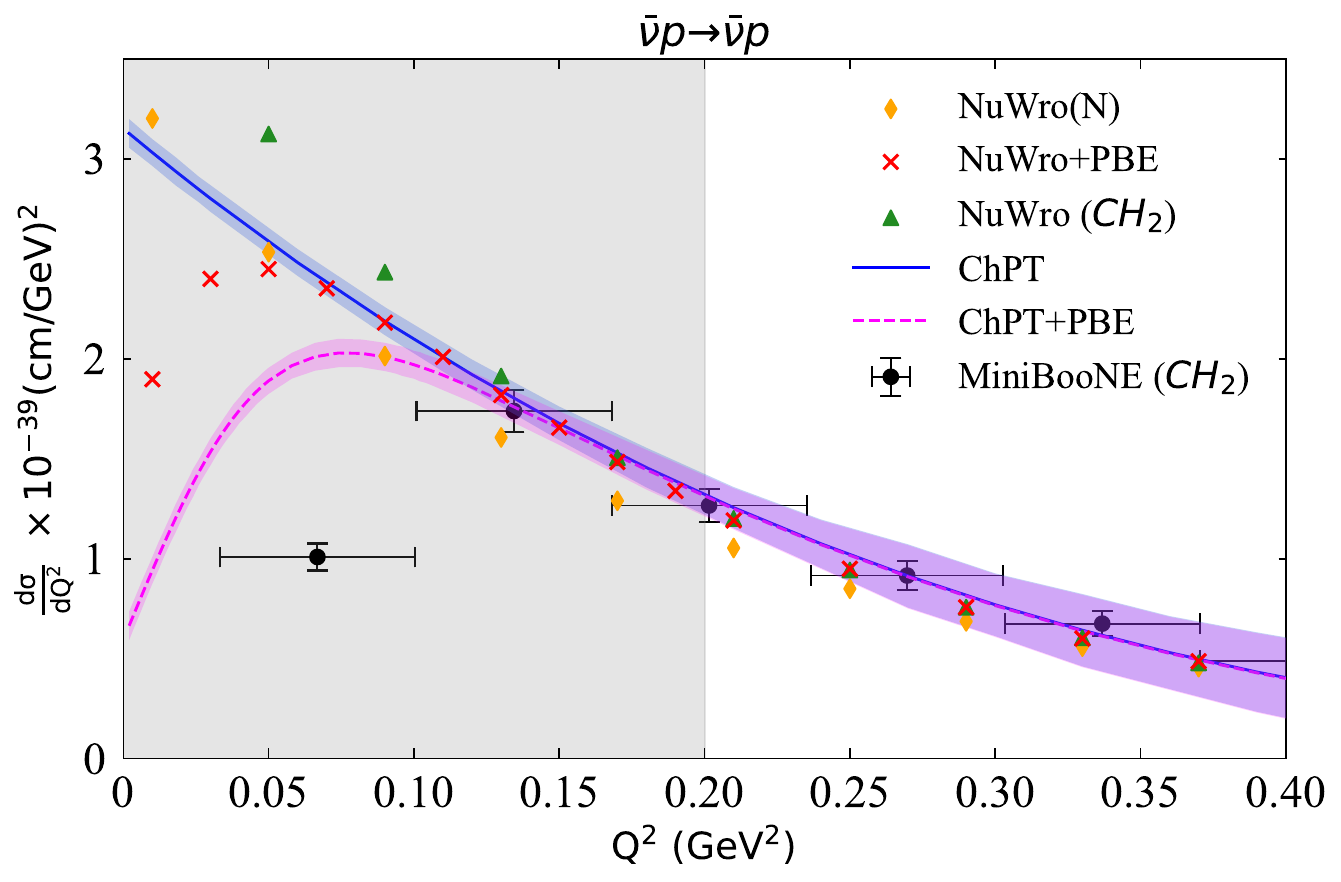} 
\caption{Differential cross section ${\rm d\sigma}/{\rm d }Q^2$ of $\bar{\nu}$-$p$ NCE scattering, with the anti neutrino energy being $E_{\bar{\nu}}=0.65$~GeV. Other description is the same as Fig.~\ref{fig:nproton_dcs_band}.} \label{fig:aproton_dcs_band}
\end{figure}

The differential scattering cross sections for $\nu$-$p$ and $\bar{\nu}$-$p$ NCE scatterings are displayed in Fig.~\ref{fig:nproton_dcs_band} and Fig.~\ref{fig:aproton_dcs_band}, respectively. Our ChPT calculation is expected to be reliable up to $Q^2=0.2$~GeV$^2$, indicated by the gray region in the figures. The first measurement of the above two processes was conducted by the BNL E734 experiment~\cite{Horstkotte:1981ne,Ahrens:1986xe}. Unfortunately, the BNL E734 data are for the $Q^2$ region higher than $0.45$~GeV$^2$, which are far beyond the validity scope of ChPT. The available experimental data for lower $Q^2$ are from the MiniBooNE experiment~\cite{MiniBooNE:2013dds,MiniBooNE:2010xqw}, which are represented by the dots with error bars in Figs.~\ref{fig:nproton_dcs_band} and~\ref{fig:aproton_dcs_band}. The MiniBooNE data for cross sections ${\rm d}\sigma/{\rm d}Q^2$ on CH$_2$ are measured at $E_\nu=0.8$~GeV for $\nu p$ and at $E_{\bar{\nu}}=0.65$~GeV for $\bar{\nu} p$. Our ChPT results are computed with the same (anti)neutrino energy for easy comparison. The blue solid lines with bands stand for the resulting ChPT predictions, which are plotted up to $E_{\nu(\bar{\nu})}=0.4$~GeV$^2$, twice of the ChPT validity limit, to see the trend of the curves. The error bands are obtained by varying the values of the LECs within their $1$-$\sigma$ uncertainties. It can be seen that our results are in good agreement with the experimental data in the range $0.13 \leq Q^2 \leq 0.2$~GeV$^2$ within uncertainties. In particular, the data at $Q^2\simeq 0.13$~GeV$^2$ are excellently described. 

However, for the $Q^2$ region lower than $\sim 0.13$~GeV$^2$, our free-nucleon scattering prediction deviates from the MiniBooNE data, which are extracted from $\nu$($\bar{\nu}$)-CH$_2$ scatterings where the nucleon are tightly bound. Such a dramatic deviation actually implies that the nuclear effects start to play an important role at low $Q^2$s. It is acknowledged that three main types of nucleon effects for bound nucleons, i.e., Fermi motion, Pauli blocking and final state interaction, may affect the cross section; see e.g. Refs.~\cite{LlewellynSmith:1971uhs,NuSTEC:2017hzk} for more details. Amongst them, the Pauli blocking effect suppresses the cross section at low momentum transfer square, as shown e.g. in Fermi gas model~\cite{Smith:1972xh}. The Pauli blocking effect has been already implemented in various neutrino events generators such as NUANCE~\cite{Casper:2002sd}, NuWro~\cite{Juszczak:2005zs,Golan:2012rfa} and so on. Here, we use NuWro to generate the Pauli blocking effect numerically, which is then added to the ChPT result. The sum of our ChPT result and the Pauli blocking effect is shown by the dashed lines with error bands in Fig.~\ref{fig:nproton_dcs_band} and Fig.~\ref{fig:aproton_dcs_band}. It can be found that the theoretical prediction is now comparable with the MiniBooNE data for the $\nu$-$p$ scattering, provided that the abnormal data point at $Q^2=0.05$~GeV$^2$ is excluded. Nevertheless, the incorporation of the Pauli blocking effect in the $\bar{\nu}$-$p$ cross section is not sufficient to make the chiral result agree with the low-$Q^2$ data at $\sim 0.067$~GeV$^2$. 

For comparison, the simulation results from NuWro event generator are also shown. The free-nucleon cross sections are marked by orange diamonds. For the cases that the target is CH$_2$ nucleus, cross sections with and without the Pauli blocking effect are represented by red crosses and green triangles. It can be observed that in the ChPT validity region, our results exhibit better agreement with the MiniBooNE data, compared to the NuWro simulation results.

\begin{figure}[tp]
\includegraphics[width=0.475\textwidth]{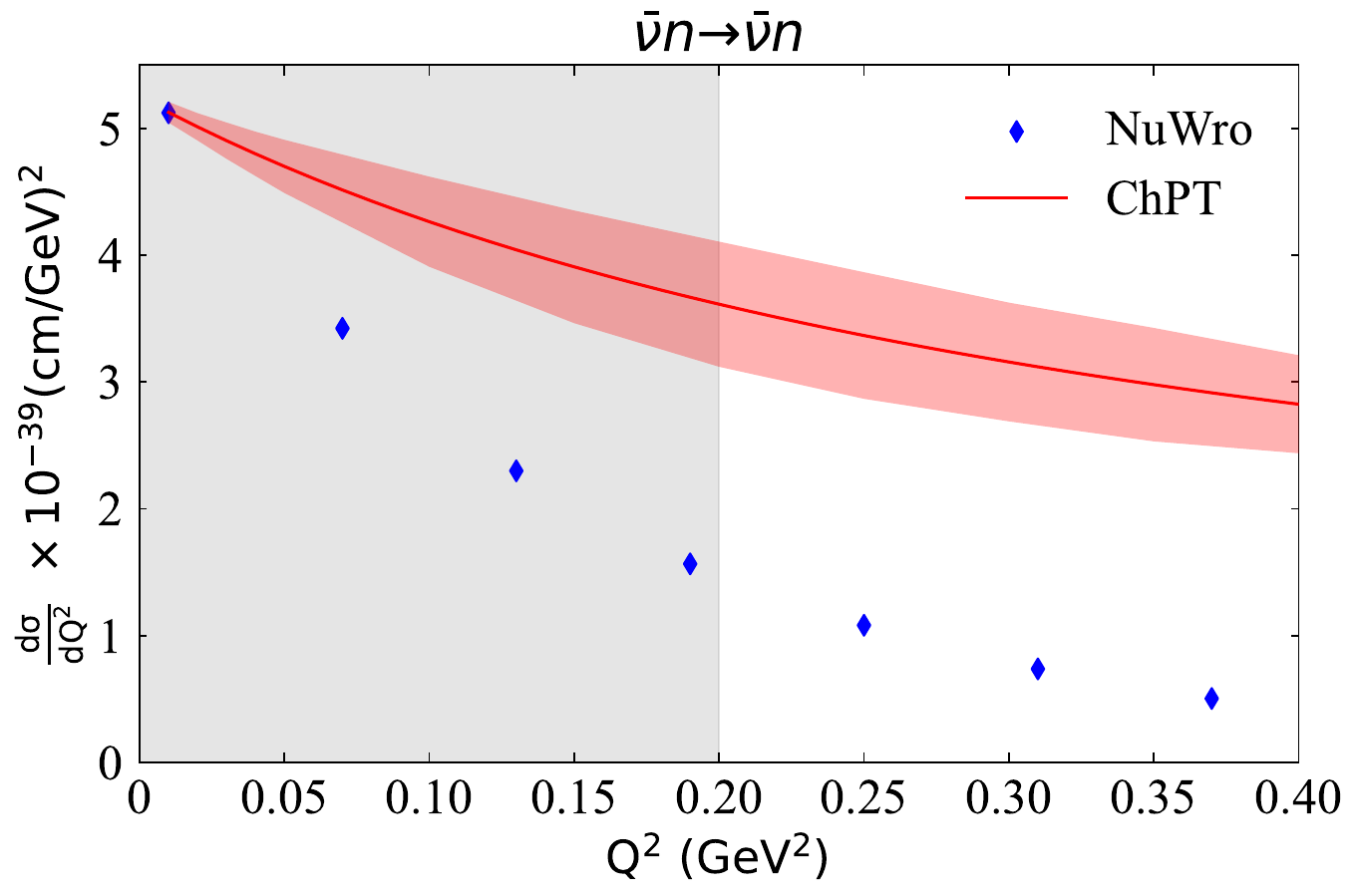}
\includegraphics[width=0.475\textwidth]{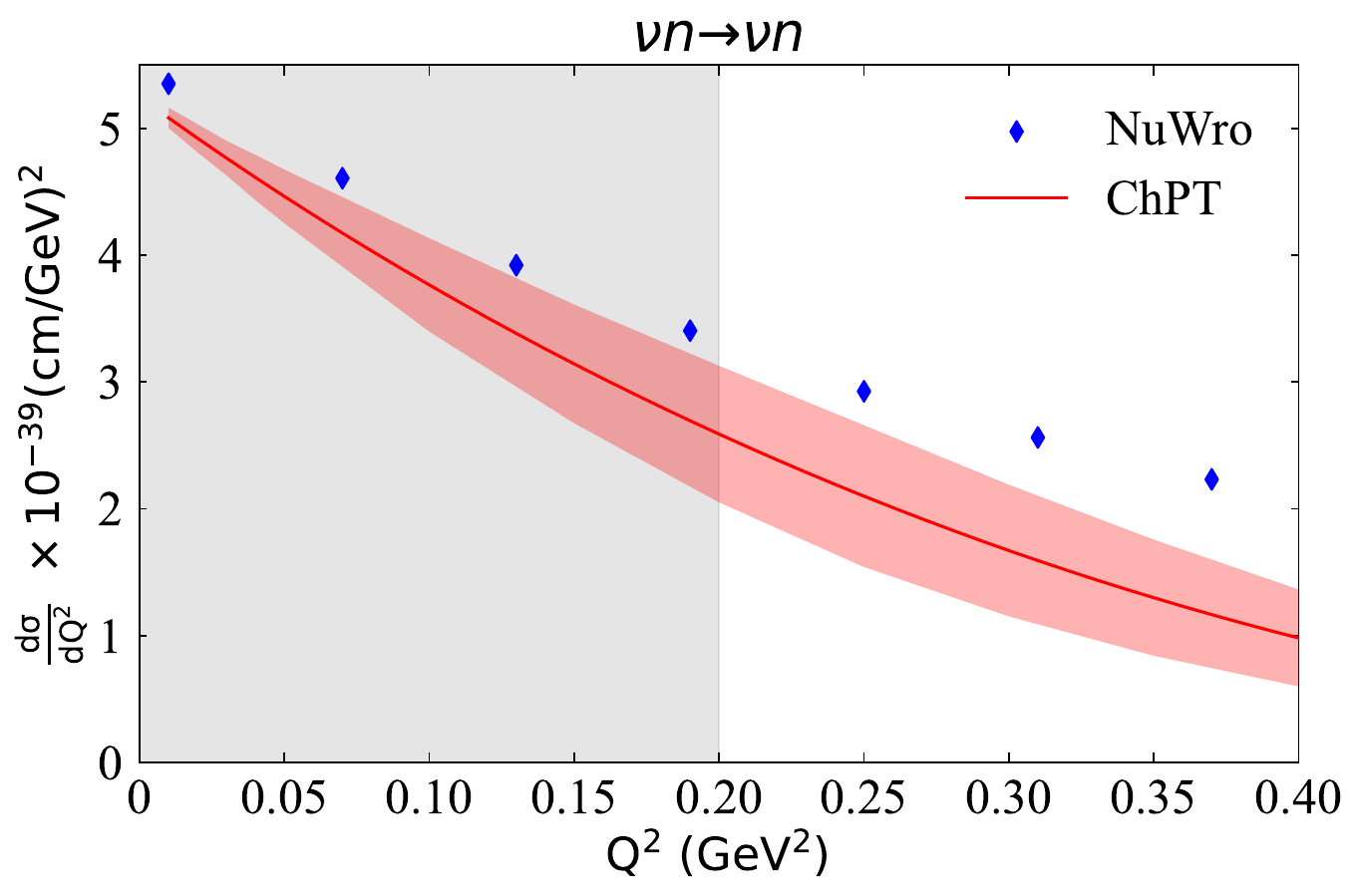}
\caption{Differential cross sections $[{\rm d}\sigma/{\rm d}Q^2]_{\nu n\to \nu n}$ with $E_\nu=0.8$~GeV and $[{\rm d}\sigma/{\rm d}Q^2]_{\bar{\nu} n\to \bar{\nu} n}$ with $E_{\bar{\nu}}=0.65$~GeV. The ChPT and NuWro simulation results are represented by solid lines and blue diamonds, respectively. The red error bands are obtained by varying the LECs in their $1$-$\sigma$ uncertainties.} \label{fig:neutron_dcs_band}
\end{figure}

Fig.~\ref{fig:neutron_dcs_band} displays the differential cross sections $[{\rm d}\sigma/{\rm d}Q^2]_{\nu n\to \nu n}$ with $E_\nu=0.8$~GeV and $[{\rm d}\sigma/{\rm d}Q^2]_{\bar{\nu} n\to \bar{\nu} n}$ with $E_{\bar{\nu}}=0.65$~GeV. Experimental data on the two processes are not available so far. For comparison, the ChPT and NuWro results are shown together. Large discrepancy between the two determinations arises for the neutron processes. Future cross section data from experiments for the neutron channels are required to explain this discrepancy, though it is a challenge due to the difficulties associated with neutron detection.

\subsection{Total Cross Section}
	
The total cross section can be readily obtained by integrating the differential cross-section over all possible scattering directions. It describes the overall probability of interaction between the incident particle and the target particle during the scattering. Taking $\nu$-$N$ NCE scattering for example, the total cross section as a function of the center-of-mass (CM) energy $\sqrt{s}$ can be written as 
\begin{equation}
\sigma ( \sqrt{s} ) = \int \frac{{\rm d} \sigma }{{\rm d} Q^2} {\rm d} Q^2 \ ,
\label{eq:sigma}
\end{equation}
where the differential cross section $ {{\rm d} \sigma }/{{\rm d} Q^2}$ is given by Eq.~\eqref{eq:dcs2}. In CM frame, the relation between the scattering angle $\theta$ and the Mandelstam variables reads
\begin{equation}
\begin{aligned}
\cos{\theta} = 1 + \frac{2 s t }{(s-m_N^2)^2}\ .
\label{eq:cos}
\end{aligned}
\end{equation}
Since $t=-Q^2$ and $s$ is a Lorentz invariant, which guarantees $s=m_N^2+2E_\nu m_N$, one can get that 
\begin{equation}
\begin{aligned}
Q^2= \frac{2m_N E_\nu^2 }{2E_\nu+m_N} ( 1 -\cos{\theta} ) , \quad  \theta \in [0, \pi] ,
\label{eq:Q2}
\end{aligned}
\end{equation}
where $E_{\nu}$ is the neutrino energy in the laboratory frame. With the above equation, it is straightforward to perform the integration in Eq.~\eqref{eq:sigma}. Moreover, the resultant total cross section is a function of $E_\nu$, i.e., $\sigma(E_{\nu})$. The above derivation also holds true for anti-neutrino nucleon NCE scattering.
	
\begin{figure}[tp]  
\centering         
\includegraphics[width=0.65\textwidth]{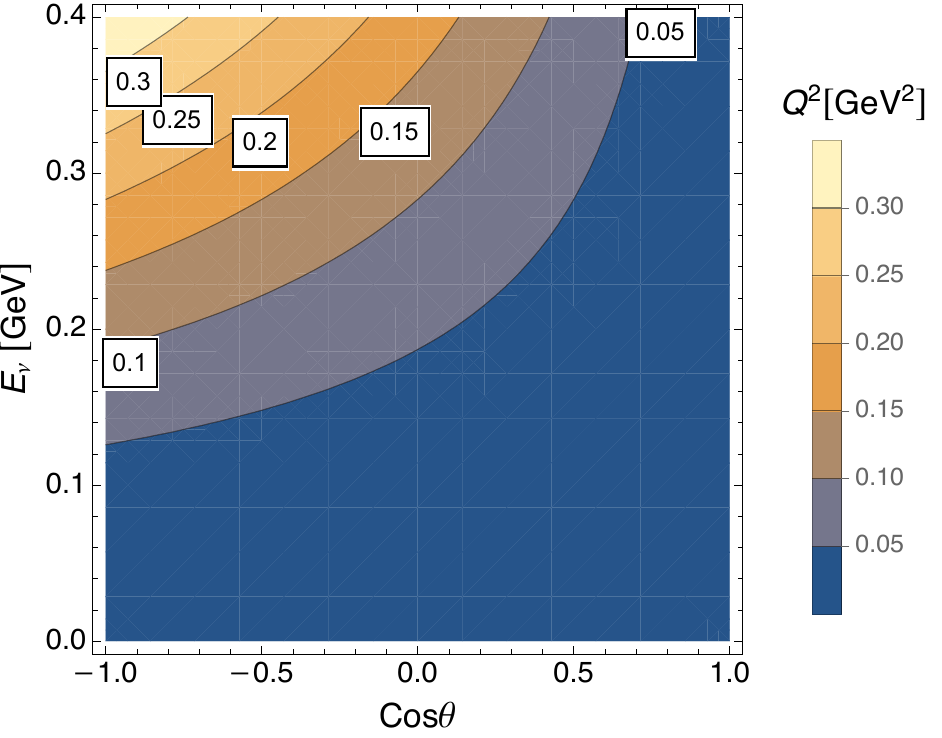}  
\caption{Contour plot of $Q^2$ with varying $E_\nu$ and $\cos\theta$.}	\label{fig:kinematics} 
\end{figure}
	
It is necessary to determine the maximal neutrino energy $E_{\nu}^{\rm max}$, up to which our ChPT model is trustworthy. For this purpose, the contour plot of $Q^2$ with respect to $E_\nu$ and $\cos\theta$ is shown in Fig.~\ref{fig:kinematics}. The ChPT validity limit of $Q^2$ is expected to be $0.2$~GeV$^2$ as mentioned in the preceding subsection. Furthermore, $\cos\theta$ must be able to take any values in the whole region $[-1,1]$, such that the numerical accuracy of the total cross section, obtained by the integration in Eq.~\eqref{eq:sigma}, is ensured. Therefore, an acceptable estimate of $E_{\nu}^{\rm max}$ is $\sim 0.28$~GeV in view of the $Q^2=0.2$~GeV$^2$ contour in Fig.~\ref{fig:kinematics}. 

\begin{figure}[tp]
\centering  
\includegraphics[width=0.96\textwidth]{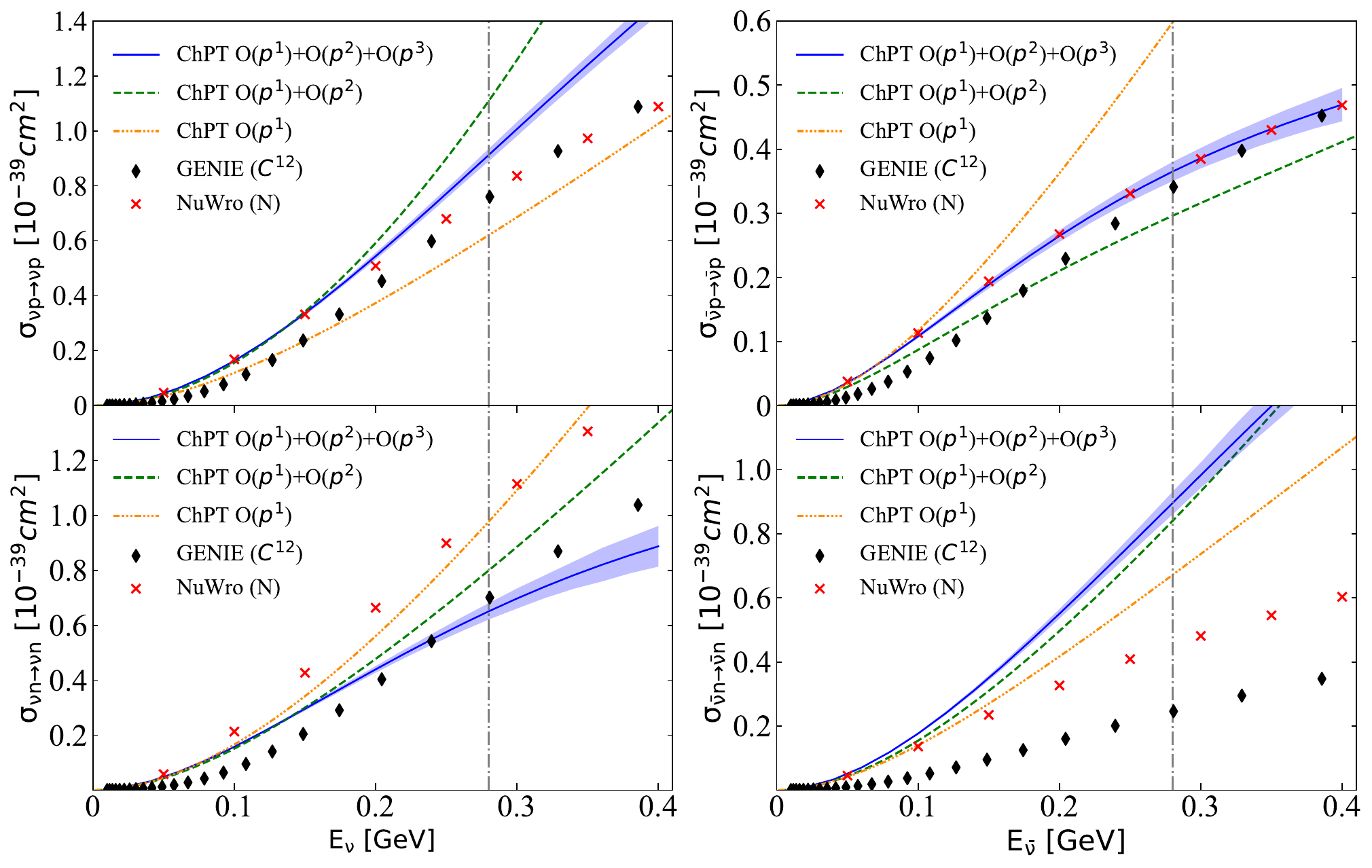}  
\caption{Total cross sections at different chiral orders. Our ChPT prediction is expected to be reliable up to $E_{\nu(\bar{\nu})}^{\rm max}=0.28$~GeV, indicated by he gray vertical line. For comparison, the simulation data produced by NuWro and GIENE events generators are also shown by red crosses and black diamonds, respectively.}	\label{fig:nuN_cs_NC}
\end{figure}

In Fig.~\ref{fig:nuN_cs_NC}, our ChPT results of the total section $\sigma (E_\nu) $ are shown order by order, together with the simulation data by NuWro~\cite{Juszczak:2005zs,Golan:2012rfa} and GENIE~\cite{Andreopoulos:2009rq}. The vertical dash-dotted lines indicate the ChPT validity limit, $E_\nu^{\rm max}\sim 0.28$~GeV, as discussed above. The NuWro data is generated for free nucleons. For the $\nu p$ and $\bar{\nu} p$ channels, a calculation with higher chiral orders makes the predictions closer to the NuWro data. The GENIE data is produced with $C^{12} $ as the target nucleus, and have been merely divided by $6$ since $C^{12} $ is composed of $6$ protons and $6$ neutrons. Namely, the nuclear effects are not excluded for the GENIE data, which leading to a sizeable deviation from our model and the NuWro data. For the $\nu n$ and $\bar{\nu} n$ channels, our ChPT predictions are consistent neither with the NuWro data nor with the GENIE data. Similar to the situation for differential cross sections, experimental data are needed to explain the deviation. 

\section{Summary and Outlook}	\label{sec4}
	
The (anti)neutrino-nucleon NCE scattering has been systematically investigated for the first time 
within the framework of covariant baryon ChPT up to $\mathcal{O}(p^3)$ in the low energy region. We have derived the model-independent hadronic amplitudes, from which explicit expressions of form factors are obtained. The UV divergences from the loops are removed by applying the $\overline{\rm MS}$-$1$ subtraction scheme, while the PCB terms are handled by using the EOMS scheme. We fix the LECs either to the values determined in previous literature or to natural size in accordance with the naturalness ansatz. Finally, numerical results of the differential cross sections and total cross sections are predicted. 

In the regime $Q^2\in[0.13,0.20]$~GeV$^2$, where nuclear effects are expected to be negligible, our ChPT predictions of $[{\rm d}\sigma/{\rm d}Q^2]_{\nu p\to\nu p}$ and $[{\rm d}\sigma/{\rm d}Q^2]_{\bar{\nu} p\to\bar{\nu} p}$ are well consistent with the MiniBooNE data within $1$-$\sigma$ uncertainties. For $Q^2\leq 0.13$~GeV$^2$, the so-call Pauli blocking effect starts to make sizeable contribution. It is found that our ChPT results can be fine tuned by incorporating the Pauli blocking effect, which is estimated by nuclear models implemented in the NuWro event generator. Moreover, our fine-tuned results are better than the simulation data, produced by NuWro, in the sense that they are more closer to the experimental data in the low $Q^2$ region. Unfortunately, there are no available experimental data for the neutron channels, due to the difficulties in neutron detection at experiments. Therefore, we prefer to compare our ChPT results of $[{\rm d}\sigma/{\rm d}Q^2]_{\nu n\to\nu n}$ and $[{\rm d}\sigma/{\rm d}Q^2]_{\bar{\nu} n\to\bar{\nu} n}$ with NuWro, but large discrepancy is observed unexpectedly. Future experimental data for these channels are needed to interpret the deviation. We also compute the total cross sections of the four physical NCE processes and confront them with the NuWro and GIENE data. Similar to the differential cross sections, our predictions of total cross sections agree with NuWro for the proton channels, but are incompatible for the neutron ones. Our ChPT calculation of the total cross section should be reliable for the (anti)neutrino energy lower than $0.28$~GeV. 
 
In the near future, low $Q^2$ data on $\nu$-$N$ NCE cross section will be available from, for instance, the MicroBooNE experiment. The model-independent ChPT results obtained in this work would be helpful for a precise determination of the strangeness axial vector form factor and hence the strangeness contribution to the nucleon spin, since they can be used in various nuclear models as inputs. On the other hand, the obtained form factors can be readily applied to investigate the CCQE scattering and muon capture processes, where the induced pseudoscalar form factor is involved. Besides, the $\Delta(1232)$ resonance contribution may be incorporated to improve the $Q^2$-behaviour of the form factors through the complexity of the $\Delta(1232)$ propagator.

\acknowledgments
This work is supported by National Nature Science Foundations of China (NSFC) under Contract Nos. 12275076, 11905258, 12335002; by the Fundamental Research Funds for the Central Universities under Contract No. 531118010379.

\appendix
\section{Definition of one-loop integrals} \label{appA}
	
In this appendix, the one-loop integrals involved in our chiral results of the hadronic amplitudes and the form factors are defined following e.g. Ref.~\cite{Denner:2005nn}.
\begin{itemize}
\item One-point one-loop integral:
\begin{align}
A_0(m^2)&\equiv \frac{\mu^{4-d}}{i} \int\frac{\mathrm{d^d}k}{(2\pi)^d}\frac{1}{k^2-m^2+i\epsilon}=-\frac{m_1^2}{16\pi^2}\bigg[R+\ln\frac{m_1^2}{\mu^2}\bigg]\ ,\label{eq.A0}
\end{align}
with $\mu$ the renormalization scale.
\item Two-point one-loop integrals:
\begin{align}
 \{B_0(b),B^\mu(b)\}&\equiv \frac{1}{i} \int\frac{\mathrm{d^d}k}{(2\pi)^d}\frac{\{1,k^\mu\}}{[k^2-m_1^2+i\epsilon][(k+p)^2-m_2^2+i\epsilon]}  \ , 
\end{align}
with the abbreviation $(b)=(p^2,m_1^2,m_2^2)$. The rank-$1$ tensor function can be further expressed as
\begin{align}
B^\mu(p^2,m_1^2,m_2^2)&=p^\mu B_1(p^2,m_1^2,m_2^2) \ 
\end{align}
The scalar coefficient function can be reduced through the Passarino-Veltmann (PV) approach~\cite{Passarino:1978jh}, which yields
\begin{align}
B_1(p^2,m_1^2,m_2^2)&=\frac{1}{2p^2}[A_0(m_1^2)-A_0(m_2^2)-(p^2-m_2^2+m_1^2)B_0(p^2,m_1^2,m_2^2)]  \ .
\end{align}
\item Three-point one-loop integrals:
\begin{align}
\{C_0(c),C^\mu(c)\} \equiv & \frac{1}{i} \int\frac{\mathrm{d^d}k}{(2\pi)^d}\frac{1}{[k^2-m_1^2+i\epsilon][(k+p_1)^2-m_2^2+i\epsilon][(k+p_2)^2-m_3^2+i\epsilon]}\ ,\notag
\end{align}
with the abbreviation $(c)=(p_1^2,q^2,p_2^2,m_1^2,m_2^2,m_3^2)$ and $q\equiv p_2-p_1$. The rank-$1$ tensor integral can be decomposed as
\begin{align}
C^\mu(p_1^2,q^2,p_2^2,m_1^2,m_2^2,m_3^2)
&=p^\mu C_1\left[p_1^2,q^2,p_2^2,m_1^2,m_2^2,m_3^2\right]\\
&+q^\mu C_2\left[p_1^2,q^2,p_2^2,m_1^2,m_2^2,m_3^2 \right]\ ,
\end{align}
where $ P\equiv p_1+p_2$. The PV reduction leads to
\begin{align}
C_1&=\frac{1}{2\big[P^2q^2-(P\cdot q)^2\big]}\bigg\{ (q^2-P\cdot q)B_0(p_1^2,m_1^2,m_2^2)+(q^2+p\cdot q)\notag\\
&\times B_0(p_2^2,m_1^2,m_3^2)
-2q^2B_0(q^2,m_2^2,m_3^2)-\Big[q^2(2m_1^2-m_2^2-m_3^2+p_1^2+p_2^2)\notag\\
&+p\cdot q(m_3^2-m_2^2-p_2^2+p_1^2)\Big]C_0(p_1^2,q^2,p_2^2,m_1^2,m_2^2,m_3^2)
\bigg\}\ ,\\
C_2&=\frac{1}{2q^2}\bigg\{ B_0(p_1^2,m_1^2,m_2^2)-B_0(p_2^2,m_1^2,m_3^2)+(m_3^2-m_2^2-p_2^2+p_1^2)
\notag\\
&\times C_0(p_1^2,q^2,p_2^2,m_1^2,m_2^2,m_3^2)
\bigg\} -\frac{p\cdot q}{q^2}C_1(p_1^2,q^2,p_2^2,m_1^2,m_2^2,m_3^2) \ .
\end{align}
\end{itemize}

\section{Hadronic amplitudes}  
\label{appB}
For brevity, we use the following abbreviations for the loop integrals:
\begin{align}
A_0^1 &= A_0(m_N^2)\ , \notag \\
A_0^2 &= A_0(m_\pi^2) \  , \notag \\
B_0^1 &= B_0(m_N^2,m_N^2,m_\pi^2)\ ,\notag  \\
B_0^2 &= B_0(t,m_N^2,m_N^2) \ , \notag \\
B_0^3 &= B_0(t,m_\pi^2,m_\pi^2) \ ,\notag \\
C_0^1 &= C_0(m_N^2,m_N^2,t,m_\pi^2,m_N^2,m_\pi^2) \ ,\notag \\
C_0^2 &= C_0(m_N^2,m_N^2,t,m_N^2,m_\pi^2,m_N^2)\ .
\label{eq:LIF}
\end{align}

In what follows, expressions of the hadronic amplitudes are listed diagram by diagram. Here, we prefer to show $\mathcal{H}^\mu$ [c.f. Eq.~\eqref{eq.GammaExp}], where the common spinors of the nucleon are thrown away for simplicity. 
\begin{itemize}
\item Diagram $(a)$:
\begin{align}
\mathcal{H}^\mu_a =&-g \frac{ \tau_3}{2}\gamma^\mu \gamma_5-\Big(  \sin^2{\theta_W}  \tau_0 - \cos{2\theta_W} \frac{ \tau_3}{2}  \Big)\gamma^\mu   \ .
\end{align}
\item Diagram $(b)$:
\begin{align}
\mathcal{H}^\mu_b =&\frac{g m_N \tau_3 }{t-M^2} q^\mu\gamma_5  \ .    
\end{align}
\end{itemize}

\begin{itemize}
\item Diagram $(c)$:
\begin{align}
\mathcal{H}^\mu_c =&-\frac{i  \sin^2{\theta_W}}{2m_N } ( c_6+2c_7 ) \sigma^{\mu \nu } q_\nu \tau_0 + \frac{i  \cos{2\theta_W}}{4m_N}c_6\sigma^{\mu \nu} q_\nu \tau_3   \ .
\end{align}
\item Diagram $(d)$:
\begin{align}
\mathcal{H}^\mu_d =&\frac{ 2 (  d_{18} -2 d_{16}  ) m_N M^2 \tau_3 }{M^2-t} q^\mu \gamma_5  \ .    
\end{align}
\item Diagram $(e)$:
\begin{align}
\mathcal{H}^\mu_e =& \frac{2 g \ell_4  m_N M^2 \tau_3 }{F^2} \frac{q^\mu \gamma_5 }{t-M^2} \ .    
\end{align}
\item Diagram $(f)$:
\begin{align}
\mathcal{H}^\mu_f=& \frac{ 2 d_7 t \sin^2{\theta_W}  \tau_0 }{m_N}   ~p^\mu    -\frac{  d_6 t \cos{2\theta_W} \tau_3  }{2m_N }  ~p^\mu -2  d_{16} M^2 \tau_3 \gamma^\mu \gamma_5  \notag  \\
& + \frac{1 }{2} d_{22} \tau_3 \Big\{ 2m_N q^\mu  - \gamma^\mu t   \Big\}  \gamma_5 \ .    
\end{align}
\item Diagram $(bs)$:
\begin{align}
\mathcal{H}^\mu_{bs}=& \frac{ 2 g m_N M^2  \Big[ M^2 \ell_3 + ( M^2 -t ) \ell_4  \Big] \tau_3 }{ F^2( M^2 -t )^2 }  q^\mu \gamma_5  \ .    
\end{align}
\item Diagram $(g)$:
\begin{align}
\mathcal{H}^\mu_g =&-\frac{\cos{(2\theta_W)} \tau_3 }{12F^2} \Big[(4m_\pi^2-t) B_0^3+2A_0^2 \Big] \gamma^\mu   \ .    
\end{align}
\item Diagram $(h)$:
\begin{align} 
\mathcal{H}^\mu_h =& ~\frac{g^2 \cos{2\theta_W} \tau_3}{12F^2 ( 4m_N^2 -t )}  \Big\{ - 24m_N^2 ( -4m_N^2 m_\pi^2 + m_\pi^4 + m_N^2 t ) C_0^1  - 24m_N^2 ( m_\pi^2 - 2m_N^2 ) B_0^1  \notag \\
& \quad - ( -40 m_N^2 m_\pi^2 +16 m_N^2 t + 4m_\pi^2 t -t^2 ) B_0^3  +  2( 4m_N^2 -t  )^2 A_0^2
\Big\}  \gamma^\mu  \notag\\
& + \frac{2 g^2 m_N \cos{2\theta_W} \tau_3}{F^2 ( 4m_N^2 -t )^2}  \Big\{ m_N^2 ( -4m_\pi^2 t - 8m_N^2 m_\pi^2 + 6m_\pi^4 + t^2 + 2m_N^2 t )  C_0^1  \notag  \\
& \quad + ( 10m_N^2 m_\pi^2 -2 m_N^2 t -4 m_N^4 -m_\pi^2 t   ) B_0^1 + 3m_N^2 ( t-2m_\pi^2 ) B_0^3 \notag \\
& \quad + ( 4m_N^2 -t  ) A_0^1  - ( 4m_N^2 -t  ) A_0^2
\Big\} ~ p^\mu    \ .    
\end{align}
\item Diagram $(i)$:
\begin{align} 
\mathcal{H}^\mu_i =&-\frac{4g m_N A_0^2  \tau_3 }{3F^2(m_\pi^2-t)} q^\mu \gamma_5  \ .    
\end{align}
\item Diagram $(j)$:
\begin{align} 
\mathcal{H}^\mu_j =&\frac{g m_N ( m_\pi^2 -4t )A_0^2 \tau_3 }{6F^2(m_\pi^2-t)^2} q^\mu \gamma_5   \ .    
\end{align}
\item Diagram $(k)$:
\begin{align} 
\mathcal{H}^\mu_k =& \frac{ \cos{2\theta_W} A_0^2 \tau_3}{2F^2} \gamma^\mu - \frac{ g  A_0^2 \tau_3}{2F^2}\gamma^\mu \gamma_5   \ .    
\end{align}
\item Diagram $(l)$:
\begin{align} 
\mathcal{H}^\mu_l =&\frac{g \tau_3}{2F^2}\left[ A_0^1 +m_\pi^2 B_0^1 \right]\gamma^\mu \gamma_5  -\frac{g^2 \cos{2\theta_W} \tau_3}{2F^2}\left[ A_0^1 +m_\pi^2 B_0^1 \right]\gamma^\mu   \ .    
\end{align}
\item Diagram $(m)$:
\begin{align} 
\mathcal{H}^\mu_m =& \frac{g \tau_3}{2F^2}\left[ A_0^1 +m_\pi^2 B_0^1 \right]\gamma^\mu \gamma_5  -\frac{g^2 \cos{2\theta_W} \tau_3}{2F^2}\left[ A_0^1 +m_\pi^2 B_0^1 \right]\gamma^\mu   \ .    
\end{align}
\item Diagram $(n)$:
\begin{align} 
\mathcal{H}^\mu_n=& ~\frac{g^3 \tau_3}{8F^2(4m_N^2-t)} \Big\{  - 2 (4m_N^2-t)  A_0^1 +  (4m_N^2-t)  A_0^2  - 2m_\pi^2(2m_N^2-t) B_0^1  \notag  \\
& \quad - 2m_N^2 ( 2m_\pi^2-t+4m_N^2 ) B_0^2  - 4m_N^2 m_\pi^4 C_0^2    \Big\}  \gamma^\mu \gamma_5  \notag  \\
	&+ \frac{g^3 m_N \tau_3 }{2F^2 t (4m_N^2 -t )} \Big\{  - 2 m_N^2 m_\pi^2 (-t + 4m_N^2-m_\pi^2 ) C_0^2  - m_\pi^2( t-2m_N^2 ) B_0^1  \notag \\
     & \quad  - m_N^2 ( -2m_\pi^2 -t +4m_N^2 )B_0^2 + (4m_N^2 -t) A_0^1 -  (4m_N^2 -t) A_0^2 \Big\}  q^\mu \gamma_5  \notag  \\
	& + \frac{g^2 \cos{2 \theta_W}\tau_3 }{8F^2( t-4m_N^2)} \Big\{  2 ( t - 4m_N^2 ) A_0^1  - ( t - 4m_N^2 ) A_0^2 + 2m_\pi^2 ( t - 6m_N^2 ) B_0^1  \notag \\
	& \quad  +  2m_N^2 ( 4m_N^2 + 2m_\pi^2 -t ) B_0^2 + 4 m_N^2 m_\pi^4 C_0^2   \Big\}  \gamma^\mu   \notag \\
	& + \frac{g^2 m_N \cos{2 \theta_W}\tau_3 }{2F^2 (t-4m_N^2)^2} \Big\{  (t-4m_N^2) A_0^1  - (t-4m_N^2) A_0^2 + m_\pi^2 ( t-10m_N^2 ) B_0^1   \notag  \\
    &  \quad + m_N^2 ( 6m_\pi^2 - t +4m_N^2 ) B_0^2 + 2 m_N^2 m_\pi^2 ( t-4m_N^2 +3m_\pi^2 ) C_0^2    \Big\}   ~ p^\mu \notag  \\   
	& + \frac{3g^2 m_N \sin^2{\theta_W} \tau_0}{F^2( t-4m_N^2 )^2}  \Big\{ ( t-4m_N^2 ) A_0^1  - ( t-4m_N^2 ) A_0^2  - m_\pi^2 ( 10m_N^2 -t ) B_0^1  \notag \\
	&  \quad - m_N^2 ( -6m_\pi^2 + t -4m_N^2 ) B_0^2  - 2 m_N^2 m_\pi^2 (  -t +4 m_N^2 -3m_\pi^2 ) C_0^2     \Big\} ~ p^\mu  \notag  \\
	& + \frac{3 g^2 \sin^2{\theta_W} \tau_0}{4F^2 ( 4m_N^2-t )}  \Big\{ 2( 4m_N^2-t ) A_0^1  -  ( 4m_N^2-t ) A_0^2  - 2 m_\pi^2 ( t- 6m_N^2 ) B_0^1   \notag  \\
	&  \quad - 2 m_N^2( 4m_N^2 +2m_\pi^2 -t ) B_0^2 - 4 m_N^2 m_\pi^4 C_0^2    \Big\}  \gamma^\mu   \ .    
\end{align}
\item Diagram $(o)$:
\begin{align} 
\mathcal{H}^\mu_o &=\frac{g m_N \tau_3}{3F^2 }\frac{A_0^2 }{t-m_\pi^2} q^\mu \gamma_5  \ .  
\end{align}
\item Diagram $(p)$:
\begin{align} 
\mathcal{H}^\mu_p &=\frac{g m_N \tau_3}{F^2 ( m_\pi^2-t ) }  \big[m_\pi^2 B_0^1 +A_0^1 \big] q^\mu\gamma_5   \ .    
\end{align}
\item Diagram $(q)$:
\begin{align} 
\mathcal{H}^\mu_q &=\frac{g m_N \tau_3}{F^2 ( m_\pi^2-t ) }  \big[m_\pi^2 B_0^1 +A_0^1 \big] q^\mu\gamma_5   \ .    
\end{align}
\item Diagram $(r)$:
\begin{align} 
\mathcal{H}^\mu_r &= \frac{g^3 m_N \tau_3}{4F^2 (t-m_\pi^2)} \Big\{  4 m_N^2 B_0^2 +  A_0^2  +4 m_N^2 m_\pi^2C_0^2 \Big\} q^\mu \gamma_5  \ .    
\end{align}
\end{itemize}

\section{Explicit expressions of form factors}\label{appC}
For easy reference, the form factors [c.f.~\eqref{eq:vas}] extracted from the hadronic amplitudes in Appendix~\ref{appB} are explicitly shown here. Abbreviations defined in Eq.~\eqref{eq:LIF} are used as well. If the result of form factor from a specific diagram is not displayed, it means that its contribution is zero.

\begin{itemize}
\item Isovector Dirac form factor: 
\begin{align}
F_1^{V,(a)}&=  1 \ ,\\
F_1^{V,(f)}&= - 2 d_6 t \ ,\\
F_1^{V,(g)}&=-\frac{ 1}{6F^2} \Big[(4m_\pi^2-t)B_0^3+2A_0^2 \Big]\ , \\       
F_1^{V,(h)}&= \frac{g^2}{6 F^2 (t-4 m_N^2)^2}  \Bigg\{ 24 m_N^2 \Big[m_N^2 (-12 m_\pi^2 t+8 m_\pi^4+3 t^2) + m_\pi^4 t\Big]C_0^1  \notag  \\
 & \quad -\Big[16 m_N^4 (8 m_\pi^2-5 t) + 4 m_N^2 t (14 m_\pi^2-5 t) + t^2 (t-4 m_\pi^2)\Big]  B_0^3  \notag \\ 
 & \quad - 24 m_N^2 \Big[ m_N^2 (6 t-16 m_\pi^2) + m_\pi^2 t\Big]B_0^1 + 48 m_N^2 (4 m_N^2-t) A_0^1 ß \notag \\
 & \quad - 2(4 m_N^2-t) (20 m_N^2+t)  A_0^2  \Bigg\}  \ , \\
F_1^{V,(k)}&=\frac{ 1}{F^2} ~ A_0^2\ , \\
F_1^{V,(l)}&=-\frac{g^2 }{F^2 }\Big[m_\pi^2 B_0^1+A_0^1 \Big] \ ,\\
F_1^{V,(m)}&=-\frac{g^2 }{F^2 }\Big[m_\pi^2 B_0^1+A_0^1 \Big]\ , \\
F_1^{V,(n)}&=- \frac{g^2}{4F^2 (t-4m_N^2)^2}  \Bigg\{ -4m_N^2m_\pi^2\Big(4m_N^2(2m_\pi^2+t)-16m_N^4+m_\pi^2t \Big) C_0^2\notag \\
&  \quad +2t(4m_N^2-t)A_0^1  +(t^2-16m_N^4)A_0^2 + 2m_\pi^2 (6m_N^2t+16m_N^4-t^2)B_0^1 \notag \\
&  \quad +2m_N^2\Big(-4m_N^2(4m_\pi^2+t)+t(t-2m_\pi^2) \Big) B_0^2   \Bigg\} \ .
\end{align}
The loop contribution is given by
\begin{align}
F^{V,loops}_1=&~ F_1^{V,(g)}+ F_1^{V,(h)}+ F_1^{V,(k)}+F_1^{V,(l)}+ F_1^{V,(m)} + F_1^{V,(n) }\ .  
\end{align}
\item Isovector Pauli form factor:
\begin{align}
F_2^{V,(c)}=&~ c_6  \ , \\
		F_2^{V,(f)}=&~2 d_6 t \ ,  \\
		F_2^{V,(h)}=&~
		\frac{8g^2 m_N^2}{F^2 \big(t-4m_N^2\big)^2} \bigg\{ m_N^2\big( 2m_N^2\big(4m_\pi^2-t\big)+4m_\pi^2t-6m_\pi^4-t^2\big)C_0^1  \notag\\
        & +3m_N^2\big(2m_\pi^2-t\big)B_0^3 + \big(2m_N^2(t-5m_\pi^2)+4m_N^4+m_\pi^2t \big)B_0^1 \notag \\
        & +\big(4m_N^2-t\big)A_0^2 -\big(4m_N^2-t\big)A_0^1  \bigg\} \ , \\
		F_2^{V,(n)} =&- \frac{2 g^2 m_N^2}{F^2 (t-4m_N^2)^2}  \Bigg\{
		m_N^2(4m_N^2+6m_\pi^2-t)B_0^2 +m_\pi^2(t-10m_N^2)B_0^1 \notag \\
        &+(4m_N^2-t)A_0^2-(4m_N^2-t)A_0^1 -2m_N^2m_\pi^2(4m_N^2-3m_\pi^2-t)C_0^2    \Bigg\} \ .
\end{align}
The loop contribution is given by
\begin{align}
F^{V,{\rm loops}}_2=& ~F_2^{V,(h)} +  F_2^{V,(n)}  \ .  
\end{align}
\end{itemize}
\begin{itemize}
\item Isoscalar Dirac form factor: 
\begin{align}
		F_1^{S,(a)}=&~ 1  \ , \\
		F_1^{S,(f)}=& ~ - 4  d_7 t  \ ,  \\		
		F_1^{S,(n)}=& ~\frac{3 g^2 }{4F^2(t-4m_N^2)^2}  \Bigg\{ -4m_N^2m_\pi^2\Big[4m_N^2(2m_\pi^2+t)-16m_N^4+m_\pi^2t\Big]C_0^2 \notag \\
		& +2m_N^2\Big[t(t-2m_\pi^2)-4m_N^2(4m_\pi^2+t)\Big]B_0^2 +(t^2-16m_N^4)A_0^2 \notag\\
		& +2m_\pi^2\Big[6m_N^2t+16m_N^4-t^2\Big]B_0^1 +2t(4m_N^2-t)A_0^1  \Bigg\} \ . 
\end{align}	
The loop contribution is given by
\begin{align}
F_1^{S,{\rm loops}}=F_1^{S,(n)}  \ .  
\end{align}
\item Isoscalar Pauli form factor: 
\begin{align}
		F_2^{S,(c)}=&~ c_6+2c_7  \ , \\
		F_2^{S,(f)}=&~ 4  d_7 t  \ , \\
	  F_2^{S,(n)}=&-\frac{6 g^2 m_N^2}{F^2(t-4m_N^2)^2} \Big\{ (4m_N^2-t)A_0^1 + (t-4m_N^2)A_0^2 +m_\pi^2\big(10m_N^2-t\big) B_0^1  \notag \\
        & +m_N^2 \big(t-4m_N^2-6 m_\pi^2\big)B_0^2 -2m_N^2m_\pi^2 \big(3m_\pi^2+t-4m_N^2 \big)C_0^2 \Big\}  \ .
\end{align}	
The loop contribution is given by
\begin{align}
  F_2^{S,loops}=F_2^{S,(n)}  \ .  
\end{align}
\end{itemize}
\begin{itemize}
\item Axial form factor: 
\begin{align}
			G_A^{(a)}=&~ g \ , \\
			G_A^{(f)}=&~ 4M^2 d_{16}+d_{22}t  \ ,  \\
			G_A^{(k)}=&~\frac{ g }{F^2} A_0^2  \ , \\
			G_A^{(l)}=&-\frac{g}{F^2} \Big[m_\pi^2 B_0^1+A_0^1 \Big]  \ , \\
			G_A^{(m)}=&-\frac{g}{F^2} \Big[m_\pi^2 B_0^1+A_0^1 \Big]  \ , \\
			G_A^{(n)}=&-\frac{g^3 }{4F^2} \frac{1}{(4m_N^2-t)} \bigg\{  -4m_N^2m_\pi^4C_0^2 +2m_N^2(-4m_N^2-2m_\pi^2+t)B_0^2  \notag \\
			&  + 2m_\pi^2(t-2m_N^2)B_0^1 + (4m_N^2-t)A_0^2 -2(4m_N^2-t)A_0^1 
			\bigg\}  \ .
\end{align}
The loop contribution is given by
\begin{align}
G^{\rm loops}_A=&~ G_A^{(k)}+ G_A^{(l)}+ G_A^{(m)}+ G_A^{(n)}   \ .  
\end{align}
\item Pseudoscalar form factor:
\begin{align}
			G_P^{(b)}=&~  \frac{2 g  m_N^2}{M^2-t}  \ , \\
			G_P^{(d)}=&~\frac{4m_N^2 M^2 (2d_{16}-d_{18})}{M^2-t}  \ , \\
			G_P^{(e)}=&~\frac{4g m_N^2 M^2 \ell_4}{F^2(M^2-t)}  \ , \\
			G_P^{(f)}=&- 2m_N^2 d_{22}  \ , \\
			G_P^{(bs)}=& -\frac{4 g m_N^2 M^2 \big[ M^2 \ell_3 + (M^2-t) \ell_4 \big]}{F^2 ( M^2-t )^2}  \ , \\
			G_P^{(i)}=&~\frac{8g m_N^2 }{3F^2(m_\pi^2-t) }A_0^2  \ , \\
			G_P^{(j)}=&-\frac{ g m_N^2 (m_\pi^2-4t) }{3F^2(m_\pi^2-t)^2 }A_0^2  \ , \\
			G_P^{(n)}=&~ \frac{  g^3  }{F^2} \frac{m_N^2}{(4m_N^2-t)t}  \Bigg\{ 2m_N^2m_\pi^2(4m_N^2-m_\pi^2-t)C_0^2  +m_\pi^2(t-2m_N^2)B_0^1  \notag  \\
			& +m_N^2(4m_N^2-2m_\pi^2-t)B_0^2 +(4m_N^2-t)A_0^2-(4m_N^2-t)A_0^1 \Bigg\}  \ , \\
			G_P^{(o)}=&~\frac{2 g m_N^2 }{3F^2(m_\pi^2-t) }A_0^2   \ , \\
			G_P^{(p)}=&-\frac{2 g m_N^2  \big[m_\pi^2 B_0^1 +A_0^1 \big]}{F^2(m_\pi^2-t)}  \ ,  \\
			G_P^{(q)}=&-\frac{2 g m_N^2  \big[m_\pi^2 B_0^1 +A_0^1 \big]}{F^2(m_\pi^2-t)}  \ ,  \\
			G_P^{(r)}=&~\frac{ g^3 m_N^2 \Big( 4 m_N^2 \big( B_0^2 + m_\pi^2C_0^2 \big) +  A_0^2 \Big)}{2F^2 (m_\pi^2-t)}   \ .
\end{align}
The loop contribution is given by
\begin{align}
G^{\rm loops}_P=&~ G_P^{(i)}+ G_P^{(j)}+G_P^{(n)} + G_P^{(o)}+G_P^{(p)}+ G_P^{(q)}+G_P^{(r)}  \ .  
\end{align}
\end{itemize}

\bibliography{references}
	
\end{document}